\documentclass[prb,twocolumn,showpacs,amsmath,amssymb,floatfix]{revtex4}

\usepackage{graphicx}
\usepackage{bm}
\usepackage{tabularx}

\newcommand{\mm}{\bm{m}}

\newcommand{\rr}{\bm{r}}
\newcommand{\m}{m^{\ast}}

\newcommand{\kk}{\bm{k}}

\newcommand{\qq}{\bm{q}}

\newcommand{\ket}{ \left. | \right\rangle}
\newcommand{\ketm}{ \left.  \right\rangle }
\newcommand{\bra}{ \left\langle | \right.}
\newcommand{\bram}{ \left\langle   \right.}

\newcommand{\se}{\mathrm{sech}}

\begin{document}

\title{Spin-wave contributions to current-induced domain wall dynamics }



\author{Yann {Le Maho}}%
\email{yann@lemaho-micro.eu}
\author{Joo-Von Kim}
\email{joo-von.kim@u-psud.fr}
\affiliation{Institut d'Electronique Fondamentale, CNRS, UMR 8622, 91405 Orsay, France}
\affiliation{Universit\'e Paris Sud, 91405 Orsay, France}
\author{Gen Tatara}
\email{tatara@phys.metro-u.ac.jp}
\affiliation{Graduate School of Science, Tokyo Metropolitan University, 1-1 Minamiosawa, Hachioji, Tokyo 192-0397, Japan}

\date{\today}

\begin{abstract}
We examine theoretically the role of spin-waves on current-induced domain wall dynamics in a ferromagnetic wire. At room temperature, we find that an interaction between the domain wall and the spin waves appears when there is a finite difference between the domain wall velocity $\dot{x}_0$ and the spin current $u$. Three important consequences of this interaction are found. Firstly, spin-wave emission leads to a Landau-type damping of the current-induced domain wall motion towards restoring the solution $\dot{x}_0 = u$, where spin angular momentum is perfectly transfered from the conduction electrons to the domain wall. Secondly, the interaction leads to a modification of the domain wall width and mass, proportional to the kinetic energy of the domain wall. Thirdly, the coupling by the electrical current between the domain wall and the spin waves leads to temperature-dependent effective wall mass. 
\end{abstract}

\pacs{75.60.Ch, 75.30.Ds, 72.25.Pn, 85.75.-d}

\maketitle

\section{\label{Introduction}Introduction}

The advent of giant magnetoresistance, magnetic tunnel junctions, and spin-transfer torque in magnetic heterostructures has led to proposals of novel applications in which magnetic domain walls are manipulated by electrical currents~\cite{Parkin04112008} instead of magnetic fields.~\cite{Allwood06142002,Ono1999} Experimental studies based on such concepts have been made possible by vast improvements in nanofabrication techniques, which allow for more precise control over domain wall nucleation and propagation.~\cite{Thomas2006,Hayashi2007} While current-driven wall motion is an important means of realizing potential applications, the threshold current for such motion still remains prohibitively high for use in integrated circuits.~\cite{Vernier2004} As a consequence, strategies are being actively sought to simultaneously achieve low threshold current densities in combination with high-speed domain wall motion.

From the point of view of fundamental physics, current-driven domain wall motion has attracted much interest because it associates a complex spin-dependent transport problem with nonlinear magnetization dynamics. This is equally true for ferromagnets based on $3d$ transition metals, such as iron, nickel, cobalt, and associated alloys, as for dilute magnetic semiconductors such as (Ga,Mn)As. From a theoretical perspective, the problem lies in computing the correct torques exerted on the magnetization by the conduction electron spins. If one assumes that the conduction electron spins, propagating with an effective drift velocity $\bm{u}$, track perfectly the local magnetization along their passage through the domain wall, one finds an additional torque on the magnetization $\bm{M}$ of the form
\begin{equation}
\bm{T}_{\rm a} = - (\bm{u} \cdot \nabla) \bm{M},
\end{equation}
which is often referred to as the ``adiabatic'' contribution of spin-transfer. This term is well-understood and has been reproduced from different transport theories.~\cite{Tatara2004,Xiao:PRB:2006,Vanhaverbeke:PRB:2007,Falloon2004,Xavier2004} The magnitude of the effective drift velocity is given by $u= j P g \mu_B/(2 e M_s)$, where $j$ is the charge current density, $P$ is the spin polarization, $\mu_B$ is the Bohr magneton, $e$ is the electronic charge, and $M_s$ is the saturation magnetization. An outstanding problem of importance concerns the origin of the so-called ``non-adiabatic'' contribution~\cite{Tserkovnyak2006}
\begin{equation}
\bm{T}_{\rm na} = \frac{\beta}{M_s} \bm{M} \times \left[ (\bm{u} \cdot \nabla) \bm{M} \right],
\end{equation}
which has been found to be necessary to describe some experimental data. $M_s$ is the saturation magnetization and the dimensionless coefficient $\beta$ characterizes the magnitude of the non-adiabatic contribution.

It is possible to gain good insight into the physics of current-driven motion through the one-dimensional model (ODM) for domain-wall dynamics. This model was much studied in the 1970s,~\cite{Schryer:JAP:1974,Malozemoff1979} and later adapted to the case of exchange torques due to coupling to conduction electrons by Berger~\cite{BERGER1978, BERGER1984, BERGER1992} and Tatara and Kohno.~\cite{Tatara2004} The ODM is derived from the Landau-Lifshitz-Gilbert (LLG) equation of motion for magnetization dynamics with the current-driven terms,\cite{Thiaville2005}
\begin{equation}
\frac{\partial \bm{M}}{\partial t} = \gamma_0 \bm{H}_{\mathrm{eff}} \times \bm{M} + \frac{\alpha}{M_s} \bm{M} \times \frac{\partial \bm{M}}{\partial t} + \bm{T}_{\rm a} +\bm{T}_{\rm na},
\label{eq:LLG}
\end{equation}
where $\gamma_0$ is the gyromagnetic constant and $\alpha$ is the Gilbert damping constant. The relevance of Gilbert damping regarding domain wall motion was dicussed by Stiles \emph{et al}.~\cite{StilesSaslow} By assuming that the domain wall shape remains rigid during propagation, it is possible to parametrize the dynamics in terms of only two conjugated coordinates: the domain wall position, $x_0$, and its conjugate momentum, $p$. By assuming that the external forces acting on the wall are sufficiently weak such that the wall shape remains rigid, the equations of motion in this limit can be written as,~\cite{Tatara2004} 
\begin{eqnarray}
\frac{dp}{dt} &=& - \frac{2 \alpha  K_{\bot} m}{ S } \frac{dx_0}{dt} +  \frac{ 2\beta K_{\bot} m }{ S } \, u + f_{\mathrm{ze}} + f_{\mathrm{pin}}; 
\label{eq:force}
\\
 \frac{dx_0}{dt} &=&  u  + \frac{\alpha S}{2 K_{\bot} m} \frac{dp}{dt} + \frac{p}{m},
\label{eq:torque}
\end{eqnarray}
where $m$ is the domain wall mass, $S$ is the spin angular momentum at each individual magnetic site and $K_{\bot}$ is the transverse anisotropy energy. The external magnetic field and a pinning potential, due to intrinsic defects or artificial pinning centers, generate the additional forces $f_{\mathrm{ze}} + f_{\mathrm{pin}}$ on the domain wall, respectively. The first equation (\ref{eq:force}) relates the domain wall acceleration $dp/dt$ to the total force. The second equation (\ref{eq:torque}) relates the domain wall velocity to the domain wall momentum. Neglecting damping, one sees that $p$ is related to the relative velocity $\dot{x}_0 - u$ through the D{\"o}ring mass of the domain wall $m = p /  ( \dot{x}_0 - u )$.

The importance of the non-adiabatic ``$\beta$-term'' is made explicit in Eq.~\ref{eq:force}, where its contribution as an effective magnetic field can be immediately seen. It has been shown in previous studies that the existence of $\beta$ leads to different qualitative dynamics for the wall motion.~\cite{Tatara2004} If $ 2 \beta \geq \sqrt{ H_p / H_{\bot} } $, with $H_p$ being the extrinsic pinning field and $H_{\bot}$ the transverse anisotropy field, the domain wall depins for $ u > \lambda \gamma_0 H_p / 2 \beta $. Therefore in the weak pinning limit, the larger is the $\beta$ term, the smaller is the critical current density.~\cite{Tatara2008}

However, as we have indicated above, the physical origin of this non-adiabatic term is still an open issue subject to spirited debate.~\cite{Beach2006} In one line of inquiry, different authors have sought to associate $\beta$ with the viscous damping coefficient $\alpha$, since both parameters describe dissipative processes.~\cite{Duine2007} Barnes and Maekawa contend that $\beta$ and $\alpha$ are equal because of Galilean invariance,~\cite{Barnes2005} while Kohno \emph{et al.},~\cite{Kohno2006} Duine \emph{et al.},~\cite{Duine2007} and Piechon and Thiaville~\cite{Piechon2006} have found that $\beta$ and $\alpha$ are not equal in general. In a different picture, Tatara and Kohno associate $\beta$ with ballistic domain wall resistance,~\cite{Tatara2004} which is independent of $\alpha$ and depends only on the transport properties of the system. Much of the difficulty in reaching a consensus is therefore related to the complexity in defining the $\beta$-term theoretically and in measuring it experimentally.

The present study is motivated by the hypothesis that the interaction between the domain-wall and spin-waves produces a term similar to the non-adiabatic term, but in the presence of \emph{only} the adiabatic component of spin-transfer. The role of spin waves on field-driven domain wall dynamics has been examined by a number of authors in the past,~\cite{Bouzidi1990,Winter1961,Janak1964} but their role on current-driven wall dynamics has not been studied in much detail theoretically. While most theories on the $\beta$-term have focused on the transport properties of the conduction electrons, few studies have considered the motion of the non-equilibrium magnetization by taking into account the fluctuations. Nevertheless, the interplay between spin waves and the domain wall should be important for at least two reasons. Firstly, thermal spin waves account for a decrease in the magnetization which can be important if the system temperature approaches the Curie temperature. This is certainly the case in dilute magnetic semiconductors. Secondly, the spin waves act as a thermal bath with which energy can be exchanged with the domain wall. Indeed, the importance of spin-waves as a channel for energy dissipation in magnetic system has long been recognized. In the context of the ferromagnetic resonance, for example, two-, three- and four-magnon processes have been shown to be crucial for explaining resonance linewidths of ferromagnetic insulators.~\cite{Sparks1964} In the context of domain wall motion, Bouzidi and Suhl have showed that power is diverted from the domain wall motion through the amplification of some thermal spin waves.~\cite{Bouzidi1990} 

Recent experimental studies suggest that current-induced domain wall motion may depend strongly on the temperature.~\cite{Laufenberg2006, Yamanouchi2006, Yamanouchi2007, Yamaguchi2007, Ravelosona2005, Dagras2007} Experiments on current-induced domain wall motion in metallic devices are generally performed at room temperature, but recently several measurements have been reported over a range of temperature from several dozens to a few hundreds of degrees Kelvin.~\cite{Laufenberg2006} Studies on the temperature are likely to bring detailed information on the current induced domain wall dynamics. The actual temperature of a ferromagnetic wire along which a charge current flows is generally modified by Joule heating, and may vary much from one sample to another depending on the efficiency by which heat is drained out. As the current density required for pushing a domain wall in a ferromagnetic metal is usually quite high, $j \approx 10^{12}$ A/m$^{2}$, the increase in the temperature due to Joule heating may even approach the Curie temperature $T_c$ \cite{Yamaguchi2007}. This could induce a drastic effect on the saturated magnetization. Similar heating effects may also appear in nanowires involving magnetic semiconductors. \cite{Yamanouchi2006,Yamanouchi2007} Laufenberger \emph{et al}. found current driven domain wall motion to be less efficient by $50$ \% when temperature increases by $200$ K.~\cite{Laufenberg2006} These authors suggest that this loss of efficiency is due to the excitation of spin waves.

In this paper, we study the role of spin waves on current-driven domain wall motion by extending the approach used by Bouzidi and Suhl,~\cite{Bouzidi1990} which associates some basic ideas form the theory of solitons~\cite{Rajaraman1982} with spin-wave theory.~\cite{White1970, Prange1979, Winter1961} The coupling between the domain wall and the thermal bath of the spin waves, which originates in the kinetic part of the spin lagrangian,~\cite{Shibata2005,Thompson2005} has a number of consequences on the current-induced domain wall motion. It leads to a new dissipation channel, whereby magnons can be absorbed or emitted as the domain-wall propagates. This dissipation channel relaxes the domain wall dynamics towards the solution $\dot{x}_0 = u$, where the domain wall velocity and the conduction electron spin current are identical and Galilean invariance is restored. This dissipation process is somewhat analogous to Landau-damping in plasmas~\cite{LandauLifshitz}. The coupling between the spin waves and the current driven domain wall also results in stochastic forces in addition to damping. These stochastic forces are weakly correlated at the time scale of the domain wall thus behave as a white noise. Besides if galilean invariance is not satisfied $u \neq \dot{x_0}$, the flow of the spin current across the wire will reduce the domain wall width and will renormalize the energy of the system. This renormalization of the system energy can be re-interpreted as a modification of the domain wall mass, which becomes temperature-dependent through the interaction with the spin-waves.

This paper is organized as follows. The spin-wave eigenmodes of the domain wall are determined in Section II. In Section III the ODM of current induced domain wall dynamics is generalized by taking the spin waves into account. In Section IV, the damping of domain wall motion through radiation of magnons is presented. This radiation leads to both $\alpha$-like and $\beta$-like terms, which are both proportional to the domain wall kinetic energy $p^2/2m$. The change in domain wall width by the electrical current is calculated in Section V and subsequently interpreted as a renormalization of the domain wall mass. The response of the spin waves to the domain wall displacement and to the spin-transfer torque is investigated in Section VI. The renormalization of the domain wall mass, as a result of this response, is then estimated numerically. In Section VII, we present some discussion and concluding remarks, as well as offering suggestions for new experiments that are designed to test the main results of our theory. The Green functions used for our calculations and the integral equation used for determining the spin-wave response are presented in the Appendix.

%
\section{Eigenmodes of a Bloch domain wall}

\label{ModesPropres}

\begin{figure}
	\begin{center}
		\includegraphics[scale=0.4]{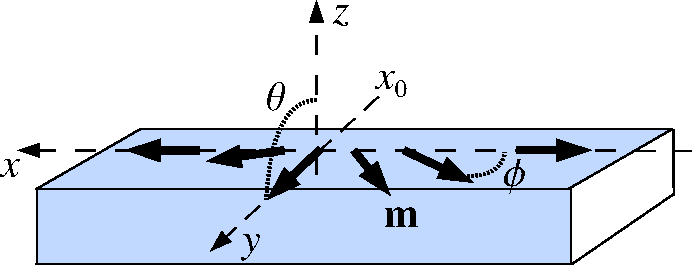}
	\end{center}
\caption{Geometry. The wire is along $x$ axis and the static domain wall profile is in the plane $(x,y)$ because of a strong perpendicular anisotropy along $z$. The spherical polar coordinates are defined with respect to the $z$ direction.}
\label{fig:figureConvention}
\end{figure}

We consider a ferromagnetic wire lying along the $x$ axis, with an axis of easy anisotropy $K_u$ along the $x$ direction and an axis of hard perpendicular (transverse) anisotropy $K_{\bot}$ along the $z$ direction (see Fig.~\ref{fig:figureConvention}). 
The orientation of the localized spins is described in spherical coordinates within a continuum approximation by means of a field $\mm(x,t)=\left( \sin \theta \cos \phi, \sin \theta \sin \phi ,\cos \theta \right)$, where $\theta(x,t)$ and $\phi(x,t)$ have a space and time dependence. In the absence of a conduction electron charge current, the magnetic energy ${\cal{H}}$ of the system is
\begin{eqnarray}
{\cal{H}} &=& \int d^3 r \left\{ A \left( ( \nabla \theta )^2 + \sin^2 \theta ( \nabla \phi )^2 \right) \right. \nonumber \\
&& \left. - K_u \sin^2 \theta \cos^2 \phi + K_{\bot} \cos^2 \theta \right\} \, ,
\end{eqnarray}
where $A$ denotes the exchange coupling. In the following we assume $ K_{\bot} \gg K_u $. This condition applies well to thin permalloy nanowires, whose in-plane (magnetocrystalline) anisotropies are very small compared to the perpendicular (demagnetizing) energy. The anisotropy constant $K_u$ describes the shape anisotropy in the plane of the wire and is as small as a few Oersteds, whereas the anisotropy constant $K_{\bot}$ describes the demagnetizing field and is about $ 4 \pi M_s \approx 10 $ kOe.

It is stated by the principle of stationary action that the energy of the magnetic system at equilibrium is minimized. In other words,
\begin{equation}
 \left. \frac{ \delta {\cal{H}}_m }{ \delta \theta } \right|_{\theta_0,\phi_0} = 0  ; \,\,\,  \left. \frac{ \delta {\cal{H}}_m }{ \delta \phi } \right|_{\theta_0,\phi_0} = 0,
\end{equation}
which leads to
\begin{eqnarray}
\theta_0 &=& \frac{ \pi }{ 2 } \, , \\
A \frac{\partial^2 \phi_0}{\partial x^2} & = &  K_u \sin \phi_0 \cos \phi_0 \, .
\end{eqnarray}
The solution is found to be a Bloch domain wall of width $\lambda= \sqrt{A/K_u}$ and energy $\sigma = 4 K_u N_{dw}$, where $N_{dw}$ denotes the number of magnetic sites inside the domain wall. More precisely the domain wall profile is given by $\sin \phi_0 = \se \left[ (x-x_0)/\lambda \right]$ and $\cos \phi_0 = - \tanh \left[ (x-x_0)/\lambda \right]$. 

\begin{figure}
  \begin{center}
          \includegraphics[scale=0.55]{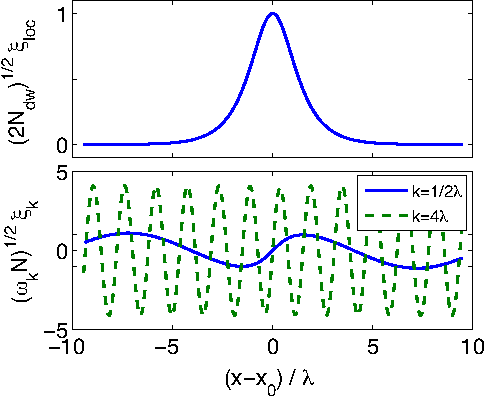}
  \end{center}
  \caption{Wavefunctions $\sqrt{ 2 N_{dw} } \xi_{\mathrm{loc}}(\rr)$ and $\sqrt{\omega_{\kk} N } \xi_{\kk}(\rr)$ about the domain wall et $x_0$, for $k_x = 1/2\lambda$ and $k_x = 4/ \lambda$. }
  \label{fig:ModeLocaliseEtEtendu2}
\end{figure}

To account for thermal fluctuations, we consider small deviations ($\delta \theta,\delta \phi$) about the static configuration ($\theta_0,\phi_0$). We expand ${\cal{H}}$ up to the second order with respect to $\delta \theta$ and $\delta \phi$ to obtain,
\begin{eqnarray}
\delta {\cal{H}} = \frac{K_u}{a^3} \int d^3 r \left\{  \delta \theta \left( {\cal{D}} + \kappa \right) \delta \theta + \delta \phi {\cal{D}} \delta \phi \right\}.
\end{eqnarray}
In agreement with earlier works,~\cite{Winter1961, Bouzidi1990} we find that the energy of the thermal fluctuations is described by a Schr\"odinger-like operator ${\cal{D}} = - \lambda^2 \partial^2_x - 2\, \se [(x-x_0)/\lambda] + 1$ with $\kappa = {K_{\bot}}/{K_u}$. 
The eigenvalues of ${\cal{D}}$ are $0$ and $\omega_{\kk} = 1 + \kk^2 \lambda^2$. The zero-eigenvalue solution $\xi_{\rm loc}$,
\begin{eqnarray}
\xi_{\mathrm{loc}}(\rr)=\frac{1}{\sqrt{2 N_{dw}}} e^{i \bm{k} \cdot \bm{r}} \se \left(\frac{x-x_0}{\lambda} \right),
\end{eqnarray}
corresponds to the Goldstone mode of the system since the energy of the static wall is independent of its position $x_0$. In other words the $\xi_{\mathrm{loc}}$-part of $\delta \phi$ contains no energy $ \xi_{\mathrm{loc}} {\cal{D}} \xi_{\mathrm{loc}} = 0 $. We can avoid expanding $\delta \phi$ on the Goldtsone mode by elevating domain wall position to a dynamical collective coordinate $x_0(t).$ \cite{Rajaraman1982} The system is not rotationally invariant about the wire axis, because of the strong perpendicular anisotropy $K_{\bot}$. As such, the $\xi_{\mathrm{loc}}$-part of $\delta \theta$ carries a finite energy and the wavefunction $\xi_{\mathrm{loc}}$ corresponds to a bound state of the system. In the following the amplitude of this bound state will be noted $c_{\mathrm{loc}}(t)$. The non-zero eigenvalues of operator ${\cal{D}}$ correspond to the propagating waves $\xi_{\kk}(\rr)$,
\begin{equation}
\xi_{\kk}(\rr) = \frac{ 1 }{ \sqrt{ \omega_{\kk} N}  } e^{i \kk \cdot \rr} \left[ \tanh{\left(\frac{x-x_0}{\lambda}\right)} - i k_x \lambda \right],
\end{equation}
we have noted $N$ the total number of magnetic sites in the sample.
 The wavefunctions $ \xi_{\kk} $ are orthonormal to each other,
\begin{equation}
\int \frac{d^3 r}{a^3} \xi_{\kk}^{\ast}  \xi_{\mm} = \delta_{\kk,\mm},
\end{equation} 
which, in turn, are orthogonal to the bound state wavefunction $ \xi_{\mathrm{loc}} $,
\begin{equation}
\int \frac{d^3 r}{a^3} \xi_{\kk}^{\ast}  \xi_{\mathrm{loc}} = 0.
\end{equation} 
The wavefunctions $\xi_{\mathrm{loc}}$ and $\xi_{\kk}$ are represented in Fig.~\ref{fig:ModeLocaliseEtEtendu2}.

It is convenient to expand the small angle deviations $\delta \theta$ and $\delta \phi$ in terms of the eigenfunctions $\xi_{\kk}$ via the complex-valued variables $d_{\kk}$ through the transformation $\delta \phi + i \delta \theta = i c_{\mathrm{loc}} \xi_{\mathrm{loc}} + \sum_{\kk} d_{\kk} \xi_{\kk}$. Using this notation, we note that the Hamiltonian can be written as ${\cal{H}} = \sigma + (\sigma/4) \sum_{\kk} \left\{  \left( \omega_{\kk}+\kappa/2 \right) {d}^{\ast}_{\kk} {d}_{\kk} - (\kappa/4) \left( {d}_{\kk} {d}_{-\kk} + {d}^{\ast}_{\kk} {d}^{\ast}_{-\kk} \right) \right\} $. This is not a quadratic Hamiltonian because of the finite perpendicular anisotropy $K_{\bot}$, which leads to elliptical spin precession. To diagonalize this Hamiltonian, we follow the usual prescription  by means of the Bogoliubov transformation ${c}_{\kk} = u_{\kk}^{+} {d}_{\kk} + u_{\kk}^{-} {d}^{\ast}_{-\kk}$, with 
\begin{equation}
u_{\kk}^{\pm} = \sqrt{ \frac{\omega_{\kk} + \kappa/2 \pm \hbar \Omega_{\kk}/K_u }{ 2 \hbar \Omega_{\kk}/K_u } },
\end{equation}
where the frequency $\Omega_{\kk}$ is defined as,
\begin{eqnarray}
\left( \frac{\hbar \Omega_{\kk}}{K_u} \right)^2 = \omega_{\kk} \left( \omega_{\kk} + \kappa \right).
\end{eqnarray}
Replacing ${d}_{\kk}$ and ${d}_{\kk}^{\ast}$ by the magnon operators ${c}_{\kk}$ and ${c}_{\kk}^{\ast}$, we get the quadratic spin-wave Hamiltonian,
\begin{eqnarray}
\delta {\cal{H}} = K_{\bot} c^2_{\mathrm{loc}} + \sum_{\kk} \hbar \Omega_{\kk} {c}^{\ast}_{\kk} {c}_{\kk} \, .
\end{eqnarray}
Spin wave energy is $\hbar \Omega_{\kk}$. The mode $c_{\kk}$ has two components $c_{\kk} = c_{\kk}^{\mathrm{def}} + c_{\kk}^{\mathrm{th}}$, which describe the wall deformation and the thermal propagating spin wave excitations. 

Next, we quantize the system by turning the complex variables $c_{\kk}^{\mathrm{th}}$ and $c^{\mathrm{th} \ast}_{\kk}$ into the boson operators $\hat{c}_{\kk}$ and $\hat{c}^{\dag}_{\kk}$, which obey the usual bosonic commutation relations. For the sake of clarity, we will find it convenient to use the variables ${\phi}_{\kk} = {c}_{\kk} + {c}^{\ast}_{-\kk} $, ${\theta}_{\kk} = (1/i) \left( {c}_{\kk} - {c}^{\ast}_{-\kk} \right) $ and their corresponding operators $\hat{\phi}_{\kk} = \hat{c}_{\kk} + \hat{c}^{\dag}_{-\kk} $, $\hat{\theta}_{\kk} = (1/i) \left( \hat{c}_{\kk} - \hat{c}^{\dag}_{-\kk} \right) $. The small angle deviations $\delta \phi$ and $\delta \theta$ are then expressed in terms of ${\phi}_{\kk}$ and ${\theta}_{\kk}$ as,
\begin{eqnarray}
\delta \phi(\rr) &=& \sum_{\kk} {\phi}_{\kk} \nu_{\kk}^{\phi} \xi_{\kk}(\rr), \label{eq:expansionTheta} \\
\delta \theta(\rr) &=& c_{\mathrm{loc}} \xi_{\mathrm{loc}}(\rr) + \sum_{\kk} {\theta}_{\kk} \nu_{\kk}^{\theta} \xi_{\kk}(\rr). \label{eq:expansionPhi}
\end{eqnarray}
The parameters $\nu_{\kk}^{\phi} = ( u_{\kk}^{+} + u_{\kk}^{-} )/{ 2 } $ and $\nu_{\kk}^{\theta} = ( u_{\kk}^{+} - u_{\kk}^{-} )/{ 2 }$ represent the ellipticity of the spin precession. If the system were rotationally invariant about the wire direction $K_{\bot} = 0$, then spin precession would be circular $ \nu_{\kk}^{\phi} = \nu_{\kk}^{\theta} = 1/2 $.

Lastly, it is convenient to renormalize the bound state amplitude $c_{loc}$ and introduce a new variable $p$ as,
\begin{eqnarray}
p = - \frac{ S \sqrt{ 2 N_{dw} } c_\mathrm{loc} }{ \lambda } \, .
\end{eqnarray} 
We will see in the next section that $p(t)$ represents the domain wall kinetic momentum.

\section{Generalized 1D model of Bloch wall dynamics}

\subsection{\label{interpretationBoundState}ODM without spin-waves}

As we will show in subsequent sections, the deformation of the domain wall due to spin transfer torques is described by both the bound state $c_\mathrm{loc}$ and the propagating states $c_{\kk}$. However in this subsection, we will disregard the spin wave modes $c_{\kk}$ and discuss the role on the domain wall dynamics of the sole bound state amplitude $c_\mathrm{loc} \equiv p(t)$. We show that we recover the usual one-dimensional model of Bloch wall dynamics without spin-waves.

In the absence of the nonadiabatic spin-transfer term, it is possible to derive the equations of motion using a Lagrangian formalism. This is particularly well-adapted to the present problem in which the magnetic system is subject to constraints related to the presence of the domain wall.

The total Lagrangian for the magnetic system is the difference between a ``kinetic'' (or Berry phase) term,
\begin{equation}
 {\cal{L}}_{\mathrm{kin}}(u=0) = S \int \frac{d^3 r}{a^3} \left( 1 - \cos \theta \right) \partial_t \phi,
\end{equation}
and the magnetic energy of the system ${\cal{H}}$. The inclusion of the adiabatic spin-transfer term appears as a moving reference frame at a velocity equal to the effective drift velocity of the spin current $u$. This is accounted for by replacing the time derivative in the kinetic part of the Lagrangian by a convective derivative,~\cite{Thiaville:JAP:2004,Shibata2005}
\begin{eqnarray}
 {\cal{L}}_{\mathrm{kin}} = S \int \frac{d^3 r}{a^3} \left( 1 - \cos \theta \right) (u\partial_x + \partial_t) \phi.
\end{eqnarray}
To zeroth order in the deformation, the kinetic part of the Lagrangian does not contribute to dynamics and can be neglected. The first-order term of ${\cal{L}}_{\mathrm{kin}}$ with respect to the deformation is
\begin{eqnarray}
{\cal{L}}^{(1)}_{\mathrm{kin}} &=& S\int \frac{d^3 r}{a^3}  c_\mathrm{loc} \xi_\mathrm{loc} ( u \partial_x + \partial_t ) \phi_0, \nonumber \\
&=& - p \left( u - \dot{x}_0 \right).
\label{eq:lagrangienordre1}
\end{eqnarray}
The overall magnetic energy is the sum of the static domain wall energy $\sigma$ and of the dipolar energy $K_{\bot} c^2_\mathrm{loc}$. The latter can be re-interpreted as the kinetic energy $p^2/2m$ of the domain wall,
\begin{equation}
{\cal{H}}_\mathrm{m} = \sigma + K_{\bot} c_{\mathrm{loc}}^2 =  \sigma + \frac{ p^2 }{ 2 m } \, , \nonumber
\end{equation}
where $m = { N_{dw} S^2 }/{ K_{\bot} \lambda^2 }$ is the D\"oring mass. Finally the full lagrangian is obtained as
\begin{equation}
{\cal{L}} = -p ( u - \dot{x}_0 ) - \sigma - p^2/2m \, .
\label{eq:lagrangienTotalRigide}
\end{equation}
Gilbert or viscous damping can be accounted for by including the dissipation function,
\begin{equation}
{\cal{F}} = \alpha S \int \frac{d^3 r}{a^3} \left(\dot{\theta}^2 + \dot{\phi}^2 \sin^2\theta  \right).
\end{equation}
Assuming the ``rigid'' domain wall profile $\theta = \theta_0(t)$ and $\phi = \phi_0(t)$,  the dissipation function ${\cal{F}}$ is readily rewritten as
\begin{equation}
{\cal{F}} = \alpha S \left( \frac{\dot{p}^2 \lambda^2}{ 2 S^2 N_{dw} } + \frac{ 2N_{dw}\dot{x}_0^2 }{ \lambda^2 }  \right).
\label{eq:fonctionDissipationGilbertRigide}
\end{equation}
The equations of motion of the domain wall coordinates $q=(x_0,p)$ are obtained by means of the Euler-Lagrange equations,
\begin{equation}
\frac{ \partial {\cal{L}} }{ \partial q } - \frac{ d }{ d t } \frac{ \partial {\cal{L}} }{ \partial \dot{q} } = \frac{1}{2} \frac{ \partial {\cal{F}} }{ \partial \dot{q} },
\label{eq:motioneulervarsigma}
\end{equation}
By combining the Lagrangian functional ${\cal{L}}$ (\ref{eq:lagrangienTotalRigide}) with the dissipation function ${\cal{F}}$ (\ref{eq:fonctionDissipationGilbertRigide}), the Euler-Lagrange equations lead to
\begin{eqnarray}
\frac{dp}{dt} + \frac{ \alpha 2 K_{\bot} }{ S } m \frac{dx_0}{dt} &=& 0 \, , \label{eq:motioneulerP} \\
\frac{ d x_0 }{ dt } - u - \frac{ \alpha S }{ m 2 K_{\bot} } \frac{dp}{dt} &=& \frac{p}{m} \, . \label{eq:motioneulerx}
\end{eqnarray}
Thus, the well-known equations of motion of the one-dimensional Bloch domain wall, (\ref{eq:force}) and (\ref{eq:torque}), are recovered. We conclude that the bound state amplitude $p(t)$ can be interpreted as the domain wall momentum as long as the propagating spin waves are neglected.

\subsection{\label{subsectionInteraction}Interaction between domain wall and propagating modes}

\begin{figure}
  	\begin{center}
  		\includegraphics[scale=0.55]{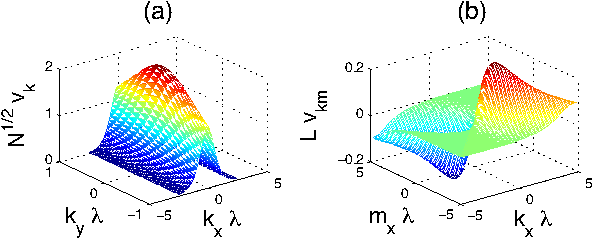}
  	\end{center}
          \caption{(a) $\sqrt{N} v_{\kk}$ as a function of $k_x$ and $k_y$. (b) $L v_{\kk\mm}$ as a function of $k_x$ and $m_x$; $K_{\bot}/K_u=57$.  }
          \label{fig:coefficientVk}
\end{figure}

The expansion of the full Lagrangian ${\cal{L}}$ up to the first order in the spin waves (\ref{eq:lagrangienordre1}) involves only the bound state and does not depend on the propagating waves. The interaction between the propagating waves and the domain wall arises from the second order expansion,
\begin{eqnarray}
{\cal{L}}^{(2)}_{\mathrm{kin}} &=&   \int \frac{d^3 r}{a^3} S \delta \theta \left( u \partial_x + \partial_t \right) \delta \phi, \nonumber \\
&=&  S \sum_{\kk} \frac{1}{4} \theta_{\kk} \dot{\phi}_{-\kk} + S( u - \dot{x}_0 ) \left[ - \sum_{\kk} v_{\kk} \frac{p}{S} {\phi}_{\kk} \right. \nonumber \\
&& \left. + \sum_{\kk \neq \mm} v_{\kk \mm}  {\theta}_{\kk} {\phi}_{\mm} + \sum_{\kk} \frac{k_x}{2} {c}^{\ast}_{\kk} {c}_{\kk} \right].
\end{eqnarray}
The coefficient $v_{\kk}$ describes a coupling between the bound state $\xi_0$ and the propagating waves,
\begin{eqnarray}
v_{\kk} = \frac{\pi}{2} \frac{1}{ \sqrt{ \omega_{\kk} N } } \nu^{\phi}_{\kk} \, \se \left( \frac{k_x \lambda \pi}{2} \right).
\end{eqnarray}
The coefficient $v_{\kk \mm}$ describes a coupling between the different propagating waves,
\begin{eqnarray}
v_{\kk \mm} &=&  \nu^{\theta}_{\kk} \nu^{\phi}_{\mm} \frac{i \pi}{2} \frac{ (k_x \lambda)^2 - (m_x \lambda)^2 }{ L \sqrt{ \omega_{\kk} \omega_{\mm} } } \times \nonumber \\
&& \mathrm{ csch } \left( \frac{ \pi \lambda ( k_x + m_x ) }{ 2 } \right) \delta_{\kk_{\|},-\mm_{\|} },
\label{eq:definitionCoefficientVkm}
\end{eqnarray}
with $ L = N_x a $ the length of the wire. Coefficients $v_{\kk}$ and $v_{\kk\mm}$ are shown on Fig.~\ref{fig:coefficientVk}. 
$v_{\kk}$ becomes negligible for $k_x > 1 / \lambda$, which means that the term $ -(u-\dot{x}_0) \sum_{\kk} v_{\kk} p \phi_{\kk} $ will mainly couple the domain wall to spin waves with a wavevector $k_x$ of the order of $1 / \lambda$. In addition, the coefficient $v_{\kk\mm}$ is large for long-wavelength spin waves $k < 1/\lambda $, $m < 1/\lambda $ and is roughly proportional to $k_x - m_x$. Thus the coupling term $ S(u-\dot{x}_0) \sum_{\kk \mm} v_{\kk \mm} \theta_{\kk} \phi_{\mm} $ will lead to a significant interaction with spin waves having $k_x \sim - m_x \sim \pm 1 / \lambda$. In some sense, the latter coupling term represents reflection of the propagating spin waves from the domain wall.

\begin{figure}
	\begin{center}
		\includegraphics[scale=0.5]{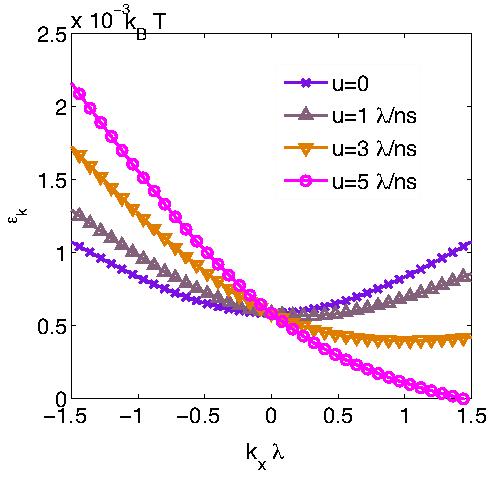}
	\end{center}
        \caption{Dispersion relation of the magnons as a function of the spin current velocity $u$, for $K_{\bot}/K_u=57$ and $\hbar / K_u = 1.88$ ns.  }
        \label{fig:PlotRelationDispersion}
\end{figure}

The quadratic term $ S( u - \dot{x}_0 ) \sum_{\kk} (k_x/2) \hat{c}^{\dag}_{\kk} \hat{c}_{\kk} $ represents a shift in the dispersion relation of the magnons. As such, the frequencies of the magnons depend on the relative velocity between the spin current and the domain wall, 
\begin{equation}
\epsilon_{\kk}  \rightarrow \epsilon_{\kk} - \frac{S}{2} k_x ( u - \dot{x}_0 ).
\label{eq:doppler}
\end{equation} 
This shift may be interpreted as a Doppler effect, which was originally put forward by Lederer and Mills~\cite{Lederer1966} and found recently in experiment by Vlaminck and Bailleul\cite{Vlaminck2008}. Some consequences of this Doppler shift have already been investigated so far, e.g. the excitation of mono-domains structures by a sole dc electrical current~\cite{Shibata2005} or as the enhancement of dissipation by spin transfer torque.~\cite{FernandezRossier2004} An example of the current-induced Doppler effect on the spin-wave dispersion relations is shown in Fig.~\ref{fig:PlotRelationDispersion}.
We note that the current-driven terms can lead to a soft-mode with a nonzero wavevector $k_c$, as indicated by the dashed line in Fig.~\ref{fig:PlotRelationDispersion} for $u=5\lambda/{\rm ns}$ for $k_c \lambda \simeq 1.5$. As pointed out by Shibata et al.,~\cite{Shibata2005} this would lead to an instability in a uniformly magnetized ground state whereby the nucleation of domains of size $\sim \pi/k_c$ is favored.

Combining the kinetic part of the Lagrangian and the magnetic energy, the full Lagrangian ${\cal{L}} $ describing the magnetic system is found to be
\begin{widetext}
\begin{equation}
{\cal{L}} = S \sum_{\kk} \frac{1}{4} \theta_{\kk} \dot{\phi}_{-\kk} +  S ( u - \dot{x}_0 )  \left[ - \frac{p}{S}  - \sum_{\kk} v_{\kk} \frac{p}{S} \phi_{\kk}   +  \sum_{\kk} \frac{k_x}{2} {c}_{\kk}^{\ast} {c}_{\kk} + \sum_{\kk \neq -\mm} v_{\kk \mm} \theta_{\kk} \phi_{\mm} \right] - \sigma -  \frac{p^2}{2m} - \sum_{\kk} \hbar \Omega_{\kk} {c}^{\ast}_{\kk} {c}_{\kk}.
\label{eq:LagrangienTotal}
\end{equation}
\end{widetext}
The classical coordinates of the domain wall are coupled to the propagating spin waves through the interaction potential,
\begin{eqnarray}
{\cal{V}} = S ( u - \dot{x}_0 ) \left[ \sum_{\kk} v_{\kk} \frac{p}{S} {\phi}_{\kk} - \sum_{\kk \mm} v_{\kk \mm} {\theta}_{\kk} {\phi}_{\mm} \right] \, .
\label{eq:interaction}
\end{eqnarray}
This potential must exist because the wavefunctions $\xi_{0}$ and $\xi_{\kk}$ don't diagonalize the kinetic part of the lagrangian ${\cal{L}}_{\mathrm{kin}}$. As such, a finite difference $u \neq \dot{x}_0$ between the spin current velocity $u$ and the domain wall velocity $\dot{x}_0$ will always give rise to a coupling between the propagating spin waves and the domain wall. However, if the spin-transfer from the adiabatic torque is achieved with an effective drift velocity such that $ u = \dot{x}_0 $, then the interaction potential ${\cal{V}}$ will vanish identically and, in that case, the domain wall and the spin waves will be completely decoupled. We point out that the solution $\dot{x}_0 = u$ is actually satisfied if the $\beta$ coefficient in the ODM is identical to the Gilbert coefficient $\alpha$, or in other words, the conservative and non-conservative dynamics of the system are invariant under a Galilean transformation.

\subsection{Force and torque}

In the following, we seek to generalize the ODM by the inclusion of the interacting potential (\ref{eq:interaction}) which couples the spin-wave modes and the domain wall. The Euler-Lagrange equations (\ref{eq:motioneulerP}) and (\ref{eq:motioneulerx}) become modified by the latter and therefore involve new terms with respect to the magnons. 

By definition, the force $F$ exerted on the domain wall is equal to the time derivative of the domain wall momentum $dp/dt$. The Euler-Lagrange equation
\begin{eqnarray}
\frac{dp}{dt} = - \frac{ \partial {\cal{H}} }{ \partial x_0 } - \frac{d}{dt} \frac{ \partial {\cal{V}} }{ \partial \dot{x}_0 }
\end{eqnarray}
indicates that $F$ originates in both the magnetic energy ${\cal{H}}$ and the interaction potential ${\cal{V}}$. 

Let us first consider the contribution $F_\mathrm{H}$ from the magnetic energy ${\cal{H}}$,
\begin{eqnarray}
F_\mathrm{H} = - \frac{ \partial {\cal{H}} }{ \partial x_0 } \, .
\end{eqnarray}
This force may be rewritten as,
\begin{eqnarray} 
 F_\mathrm{H} = - 4 K_u \int \frac{d^3 r}{a^3} \delta m_z(x) \partial_x \mathrm{sech}^2(x-x_0) \, ,
\label{eq:forcem}
\end{eqnarray}
where $\delta m_z$ represents the decrease in the longitudinal component of the magnetization $\delta m_z = \left( \delta \theta^2 + \delta \phi^2 \right) / 2  $ and is finite at finite temperatures due to thermal exictations. By inspection of (\ref{eq:forcem}), we observe that $F_\mathrm{H}$ is only finite if $\delta m_z$ is odd with respect to the domain wall position $\delta m_z(x_0 - x_1) = - \delta m_z(x_0 + x_1)$. This can be the case if the an electrical current flows through the domain wall and breaks the symmetry of the system. 

Let us now consider the force $- \frac{d}{dt} \frac{d {\cal{V}} }{d \dot{x}_0 }$ coming from the interaction potential ${\cal{V}}$. This force may be divided into three different components: $F_\mathrm{j}$, $F_\mathrm{def}$ and $F_\mathrm{stoc}$. The force $F_\mathrm{j}$,
\begin{eqnarray}
F_\mathrm{j} =  S  \frac{d}{dt} \sum_{\kk \mm} v_{\kk \mm} \bram \hat{\theta}_{\kk}  \hat{\phi}_{\mm} \ketm  \, ,
\label{eq:forcej}
\end{eqnarray}
originates from the coupling between the different spin-wave modes induced by the electrical current. Because $\bram \hat{\theta}_{\kk}  \hat{\phi}_{\mm} \ketm$ depends on the statistics of the magnons, the force $F_\mathrm{j}$ is expected to depend on the temperature. The force $F_\mathrm{def}$ is related to the deformation modes $c_{\kk}^{\mathrm{def}}$ (see \ref{ModesPropres}). By writing $\phi_{\kk}^{\mathrm{def}} = c_{\kk}^{\mathrm{def}} + c_{-\kk}^{\mathrm{def}  \ast} $, we find
\begin{eqnarray}
F_\mathrm{def} = -  \frac{ d }{ dt } \sum_{\kk} v_{\kk} p \phi_{\kk}^{\mathrm{def}} \, . 
\label{eq:DefinitionForceDeformation}
\end{eqnarray}
In contrast to $F_\mathrm{j}$ and $F_{\mathrm{def}}$, the force $F_{\mathrm{stoc}}$ is not deterministic but stochastic. Its average value vanishes $\bram F_{\mathrm{stoc}} \ketm = 0$, but its autocorrelation function is finite $ \bram F_{\mathrm{stoc}}(t) F_{\mathrm{stoc}}(t') \ketm \neq 0 $. The force $F_{\mathrm{stoc}}(t)$ may be rewritten as,
\begin{eqnarray}
F_{\mathrm{stoc}} = -  \frac{ d }{ dt } \sum_{\kk} v_{\kk} p \hat{\phi}_{\kk} \, .
\label{eq:expressionForceStochastique}
\end{eqnarray} 
Adding all these forces together, the total force $F$ exerted on the domain wall is found as
\begin{eqnarray}
\frac{dp}{dt} =  F_\mathrm{H} + F_\mathrm{j} + F_{\mathrm{def}} + F_{\mathrm{stoc}} 
\label{eq:pfd}
\end{eqnarray}
It will be shown in the following that all the forces $F_\mathrm{H}$, $F_\mathrm{j}$ and $F_{\mathrm{def}}$ can be re-interpreted as a modification of the domain wall mass.

So far we have looked at the contribution of the spin waves to the force $F$ but still not to the torque $T$. The torque $T$ is derived by taking the Euler-Lagrange equations with respect to the momentum $p$,
\begin{eqnarray}
\dot{x}_0  = \frac{p}{m} + u - \sum_{\kk} v_{\kk} \phi_{\kk} (\dot{x}_0-u).
\label{eq:torqueStochastique}
\end{eqnarray}
In addition to the spin-transfer drift velocity $u$ and to $p/m$, which represents the torque due to the demagnetizing field, the equation (\ref{eq:torqueStochastique}) involves a stochastic torque $T_{\mathrm{stoc}}$,
\begin{eqnarray}
T_{\mathrm{stoc}} = ( \dot{x}_0 - u ) \sum_{\kk} v_{\kk} \hat{\phi}_{\kk}
\label{eq:expressionCoupleStochastique}
\end{eqnarray}
This torque induced by the spin waves, which increases as a function of the relative velocity $ \dot{x}_0 - u $.

\section{Spin-wave emission}

\subsection{\label{DissipationFunction}Dissipation function}

According to (\ref{eq:expressionForceStochastique}) and (\ref{eq:expressionCoupleStochastique}), spin waves exert a non deterministic force and torque on the domain walls. This is because the domain walls are sensitive to the fluctuations of the spin waves through the interaction potential
\begin{eqnarray}
{\cal{V}}_{\mathrm{int}} = (u-\dot{x}_0) p \sum_{\kk} v_{\kk} \left( \hat{c}_{\kk} + \hat{c}^{\dag}_{-\kk} \right) \, .
\end{eqnarray}
We know by the fluctuation-dissipation theorem that these fluctuations necessarily lead to dissipation. As we haven't taken this dissipation into account yet, our ODM with the forces $F_\mathrm{H}$, $F_\mathrm{j}$, $F_\mathrm{def}$, $F_\mathrm{stoc}$ and the torque $T_\mathrm{stoc}$ is not fully consistent.

The spin current supplies energy to the domain wall motion by increasing the kinetic energy $ { p^2 }/{ 2 m } $. However a part of this energy is lost by the domain wall and is transfered to the lattice through a series of relaxation processes. The transfer of energy to the lattice accompagnying a domain wall motion is due to various channels of relaxation\cite{Prokof'ev2000}. One channel of relaxation starts with the emission of magnons by the excited domain wall. 

Caldeira and Leggett,~\cite{Caldeira:1983} following on from the seminal work and Feynman and Vernon,~\cite{Feynman:AP:1963} have shown irreversibility to arise when a moving particle is coupled to numerous degrees of freedom. In our case, the magnetic domain wall represents the moving particle, which propagates through a dissipative environment represented by the thermal spin-waves. The domain wall is described by its position $x_0$ and its kinetic momentum $p$, whereas the spin wave environment is described by a set of oscillators,
\begin{eqnarray}
{\cal{H}}_{\mathrm{bath}} = \sum_{\kk} \hbar \Omega_{\kk} \hat{c}^{\dag}_{\kk} \hat{c}_{\kk}.
\end{eqnarray}
The bath of magnons is assumed to be always at thermal equilibrium which is reached through very fast three and four magnon processes and due to the interaction with phonons. The domain wall and the spin wave environment are coupled to each other through the ``linear'' interaction potential ${\cal{V}}_\mathrm{int}$. Notice the similarity between the interaction ${\cal{V}}_\mathrm{int}$ and the interaction between electrons and phonons in the metals. The present system is actually formally equivalent to the dissipative system in Caldeira and Leggett's general theory, which itself is in agreement with the fluctuation dissipation theorem.~\cite{Fujikawa1998,Fujikawa1998a} Notice that a coupling very close to ${\cal{V}}_\mathrm{int}$ has already been considered by Thomson and Stamp, who investigated the damped motion of vortices driven by a magnetic field.~\cite{Thompson2005}

The emission of a magnon $\kk$ is represented by the term,
\begin{eqnarray}
{\cal{V}}_\mathrm{em}^{\kk} &=& ( u -\dot{x}_0 ) p v_{\kk} \hat{c}^{\dag}_{\kk} \, .
\end{eqnarray}
Similarly the absorption by the domain wall of a magnon $\kk$ is represented by,
\begin{eqnarray}
{\cal{V}}_\mathrm{ab}^{\kk} &=& ( u -\dot{x}_0 ) p v_{\kk} \hat{c}_{\kk} \, .
\end{eqnarray}
The energy transfer from the domain wall to the spin waves ensemble will correspond to the difference between the emission of magnons and the absorption of magnons.

In the following, the various states of the spin wave ensemble will be denoted by $ | E_n \ketm $ and the density of state of the spin waves will be represented by $\rho(E)$. By Fermi's golden rule, if the spin waves ensemble were in the state $ | E_n \ketm $, then the exchange of a magnon $\kk$ between the domain wall and the bath of the spin waves would statistically decrease the domain wall's energy by an amount $F_n(\kk)$ as,
\begin{eqnarray} 
F_n(\kk) &=& 2 \pi \Omega_{\kk} \nonumber \\
&& \left[ \sum_{ m } | \bram E_{m} | {\cal{V}}_{em}^{\kk} | E_n \ketm |^2 \delta( E_{m} - E_n - \hbar \Omega_{\kk} ) \right. \nonumber \\
&& \left. - \; \sum_{ m } | \bram E_{m} | {\cal{V}}_{ab}^{\kk} | E_n \ketm |^2 \delta( E_{m} - E_n + \hbar \Omega_{\kk} )  \right], \nonumber
\end{eqnarray}
which can be written explicitly as,
\begin{eqnarray} 
F_n(\kk) &=& 2 \pi \Omega_{\kk} \left[ \sum_{ m } | ( u -\dot{x}_0 ) p v_{\kk} |^2 \delta( E_{m} - E_n - \epsilon_{\kk} ) \right. \nonumber \\
&& \left. - \; \sum_{ m } | ( u -\dot{x}_0 ) p v_{\kk} |^2 \delta( E_{m} - E_n + \epsilon_{\kk} )  \right].
\end{eqnarray}
In reality the probability for the spin wave ensemble to remain in the state $ \bram E_n \ketm $ is equal to the Boltzmann factor $P_n = \exp \left( - \beta E_n \right) $. The magnons $\kk$ thus contribute to a transfer of energy from the domain wall to the spin waves ensemble by an amount $ F(\kk) = \sum_{n} P_n F_n(\kk) $. In the continuum limit $\sum_n E_n \rightarrow \int dE$,
\begin{eqnarray}
F(\kk) = \int_0^{\infty} dE \rho(E) f(E) F_E(\kk) \, ,
\end{eqnarray}
where $f(E) = \exp{( - \beta E )} $ and $F_E(\kk)$ is,
\begin{eqnarray} 
F_E(\kk) &=& 2 \pi \Omega_{\kk} | ( u -\dot{x}_0 ) p v_{\kk} |^2 \times \nonumber \\
&& \sum_m \left[  \delta( E_{m} - E - \hbar \Omega_{\kk} ) - \delta( E_{m} - E + \hbar \Omega_{\kk} )  \right] \, .\nonumber
\end{eqnarray}
In other words the energy lost by the domain wall through the transfer of a magnon $\kk$ is,
\begin{eqnarray}
F(\kk) =  | ( u -\dot{x}_0 ) p  |^2 \Omega_{\kk}^2 v_{\kk}^2 R ( \kk ) \, ,
\end{eqnarray}
where the function $R ( \kk )$ is closely related to the spectral function of the environment and equals
\begin{eqnarray}
R ( \kk ) = \int_0^{\infty} dE \rho(E) f(E) \frac{2 \pi}{\Omega_{\kk}}  \left[  \rho( E + \epsilon_{\kk} ) - \rho( E - \epsilon_{\kk} )  \right] \, , \nonumber
\end{eqnarray}
with $ \rho(E) $ being the density of states $\rho(E) = \sum_m \delta(E_m - E)$ of the spin wave ensemble. The total energy lost by the domain wall per unit of time $ {\cal{F}} = \sum_{\kk} F_{\kk} $ is finally obtained as,
\begin{eqnarray}
{\cal{F}}  = \sum_{\kk} | ( u -\dot{x}_0 ) p  |^2 \Omega_{\kk}^2 v_{\kk}^2 R ( \kk ) \, .
\end{eqnarray}
Let us introduce the dimensionless friction paramter $\eta =  \sum_{\kk} \Omega_{\kk}^2 v_{\kk}^2 R ( \kk ) $. The dissipation function can then be rewritten as,
\begin{eqnarray}
{\cal{F}} = \frac{2 \eta }{ N_{dw} S } | ( u -\dot{x}_0 ) p |^2  \, .
\end{eqnarray}

\subsection{$\beta$ term}

To understand how this dissipation function affects current-induced domain wall motion, it is convenient to perform a Galilean transformation and use the local frame moving at the velocity of the spin current. The domain wall position in this frame is
\begin{equation}
q = x_0 - u t.
\end{equation}
The dissipation function $ {\cal{F}} = - d {\cal{H}} / dt  $ can be rewritten in terms of $q$ as,
\begin{eqnarray}
{\cal{F}} = \frac{2 \eta }{ N_{dw} S } | \dot{q} p |^2 
\label{eq:expressionFonctionDissipation}
\end{eqnarray} 
thus treating $q$ and $p$ as independent variables,
\begin{eqnarray}
{\cal{F}} = \frac{1}{2} \frac{ \partial {\cal{F}} }{ \partial {\dot{q}} } \dot{q}
\end{eqnarray}
If the system is translationally invariant (no extrinsic pinning), the lagrangian describing the domain wall motion will not depend on the the domain wall position $q$
\begin{eqnarray}
{\cal{L}} = - {\cal{H}} + \frac{ \partial {\cal{L}} }{ \partial \dot{q} } \dot{q}
\end{eqnarray}
 therefore the time derivative of the domain wall Hamiltonian becomes,
\begin{eqnarray}
\frac{ d{\cal{H}} }{dt} = \left( \frac{d}{dt} \frac{ \partial {\cal{L}}  }{ \partial \dot{q} } - \frac{ \partial {\cal{L}} }{ \partial q } \right) \dot{q} \, .
\label{eq:equationDeLaDynamique}
\end{eqnarray}
Combining (\ref{eq:equationDeLaDynamique}) with the definition of the dissipation function ${\cal{F}}$, we get
\begin{eqnarray}
\frac{d}{dt}   \frac{ \partial {\cal{L}} }{ \partial \dot{q} } - \frac{ \partial {\cal{L}} }{ \partial q } + \frac{1}{2} \frac{ \partial {\cal{F}} }{ \partial \dot{q} } = 0 \, ,
\label{eq:generalizationmodel1D}
\end{eqnarray}
which after carrying out (\ref{eq:lagrangienTotalRigide}) and (\ref{eq:expressionFonctionDissipation}) yields,
\begin{eqnarray}
\dot{p} + \frac{ 2 \eta p^2 }{ N_{dw} S } \dot{x}_0 = \frac{ 2 \eta p^2 }{ N_{dw} S } u \, .
\label{eq:mouvementDissipationSW}
\end{eqnarray}
The dissipation by spin wave emission gives rise to a damping term $ 2 \eta p^2 \dot{x}_0 / N_{dw} $ proportional to the domain wall velocity $\dot{x}_0$ and to a force $ 2 \eta p^2 u / N_{dw} S$ proportional to the spin current $u$. These are equivalent to a damping coefficient $\alpha_{sw}$ and to a coefficient $\beta_{sw}$ expressed by,
\begin{eqnarray}
\alpha_{sw} = \beta_{sw} = \frac{ \eta p^2 }{ N_{dw} K_{\bot} m } \, .
\label{eq:valeurDeAlphaEtBeta}
\end{eqnarray}  
It is seen that the spin wave contributions $\alpha_{sw}$ and $\beta_{sw}$ are identical to each other and are also proportional to the domain wall kinetic energy $p^2/2m$. 

The dissipation function generally tends to decrease during the motion, because it vanishes at equilibrium. In the case of Gilbert damping, which is described by the dissipation function $ F/2SN_{dw} = \alpha (\dot{x}_0/\lambda)^2 (1+\alpha^2) $, dissipation thus tends to reduce the domain wall velocity $\dot{x}_0$ to zero. In contrast, the spin waves which lead to the dissipation function (\ref{eq:expressionFonctionDissipation}) will not tend to lower the domain velocity $\dot{x}_0$, but instead the relative velocity $ \dot{q} = \left| \dot{x}_0 - u \right| $. In other words, whereas Gilbert damping slows down the mowing domain walls and reduce their velocity to zero, the damping by the spin waves acts towards restoring the solution $\dot{x}_0 = u$. 

The equality between $\alpha$ and $\beta$ restores Galilean invariance, which has previously been argued by Barnes and Maekawa.~\cite{Barnes2005} In our theory, this invariance is restored because the dissipation channel, in this case the magnons, also ``flows'' with the effective drift velocity $u$ through the Doppler effect. As such, the dissipation of the domain wall motion through spin-wave interactions leads to a Landau-type damping in which the motion tends towards $\dot{x}_0 = u$. We contend, therefore, that \emph{any} dissipation channel that drifts at the same velocity as the spin-current $u$ should lead to a similar dissipation channel, for which we can write symbolically as $\alpha_{\rm drift} = \beta_{\rm drift}$.

Of couse, extrinsic pinning centers such as magnetic impurities break the translational invariance of the wire, so (\ref{eq:mouvementDissipationSW}) would no longer apply and the full coefficients $\alpha$ and $\beta$ should not be equal in general. Notice that the damping coefficient $\alpha_{sw}$, which we have identified in (\ref{eq:valeurDeAlphaEtBeta}), doesn't exactly recover the Gilbert coefficient $\alpha$, since it doesn't appear in the Euler Lagrange equation with respect to $p$.

\subsection{Stochastic field} 

The force $F_\mathrm{stoc}$ (\ref{eq:expressionForceStochastique}) and the torque $T_\mathrm{stoc}$ (\ref{eq:expressionCoupleStochastique}), which accompany the dissipation by the magnons, lead to a domain wall motion that is stochastic. As $\bram \hat{\phi}_{\kk} \ketm = 0$ and $\bram \hat{\theta}_{\kk} \ketm = 0$, the average values of $F_{\mathrm{stoc}}(t)$ and $T_{\mathrm{stoc}}(t)$ vanish identically, but their correlations are finite. Assuming the wire to be very large compared to the domain wall width $\lambda$, the auto-correlation of the force $F_\mathrm{stoc}$ is found to be
\begin{eqnarray}
\bram F_\mathrm{stoc}(t) F_\mathrm{stoc}(t') \ketm = \dot{p}(t) \dot{p}(t') \left( \frac{a}{2\pi \lambda} \right)^3 \frac{ 2 k_B T }{ K_u } I(t-t') \nonumber \\
 + \; p(t) p(t') \left( \frac{K_u}{\hbar} \right)^2 \left( \frac{a}{2\pi \lambda}  \right)^3 \frac{ 2 k_B T }{ K_u } J(t-t') \, , \nonumber
\end{eqnarray}
\begin{figure}
  	\begin{center}
  		\includegraphics[scale=0.5]{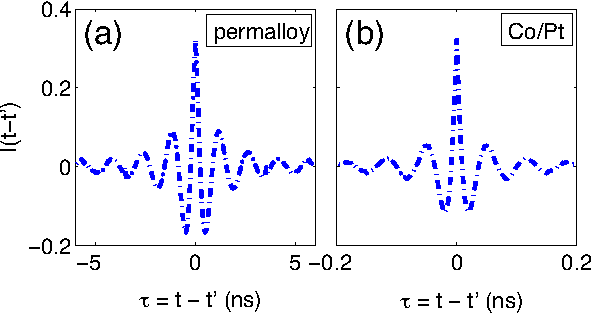}
  	\end{center}
          \caption{(a) Correlation function $I(t-t')$ between the stochastic torque $T_\mathrm{stoc}(t)$ at two different times $t$ and $t'$, for a wire of Permalloy~\cite{Thiaville2005}, with $K_{\bot}/K_u=57$ and $\hbar /K_u=1.88$ ns. (b) $I(t-t')$ for a wire of Co/Pt~\cite{Cayssol2004}, with $K_{\bot}/K_u=0.73$ and $\hbar/K_u=0.013$ ns. }
          \label{fig:IPermalloy}
\end{figure}
with,
\begin{eqnarray}
I(t-t') = \int d^3 k \lambda^3 (\eta^{\phi}_{\kk})^2 \left( \frac{\pi}{2} \right)^2 \mathrm{sech}^2 \left( \frac{ k_x \lambda \pi }{ 2 } \right)  \nonumber \\ 
\times \; \frac{ \cos \Omega_{\kk}(t-t') }{ \omega_{\kk}^{\frac{3}{2}} (\omega_{\kk}+\kappa)^{\frac{1}{2}} },
\label{eq:fonctionI}
\end{eqnarray}
and
\begin{eqnarray}
J(t-t') = \int d^3 k \lambda^3 (\eta^{\phi}_{\kk})^2 \left( \frac{\pi}{2} \right)^2 \mathrm{sech}^2 \left( \frac{ k_x \lambda \pi }{ 2 } \right)  \nonumber \\ 
\times \; \sqrt{ \frac{ \omega_{\kk} + \kappa }{ \omega_{\kk} } } \cos \Omega_{\kk} (t-t').
\end{eqnarray}
$I(t-t')$ is presented on Fig.~\ref{fig:IPermalloy} with the anisotropy constants $K_u$ and $K_{\bot}$ taken to be those for permalloy \cite{Thiaville2005} and Co/Pt \cite{Cayssol2004}. It is seen that $I(t-t')$ cancels out very quickly when $|t-t'|$ becomes larger than 1 ns. Besides the function $J(t-t')$ behaves very similarly and weakens very quickly when the time difference $|t-t'|$ increases. We therefore conclude that the stochastic forces $F_\mathrm{stoc}(t)$ and $F_\mathrm{stoc}(t')$ are not correlated to each other at the characteristic time scale associated with domain wall motion, which usually is longer than the nanosecond.~\cite{Klaui2005} For instance, the timescale of a domain wall at a typical velocity of 1 m/s in a permalloy wire corresponds to $\lambda / 1m.s^{-1} \sim$ 50 ns.

Analogously, the auto-correlation of the torque $T_\mathrm{stoc}(t)$ is found to be
\[
\bram T_\mathrm{stoc}(t) T_\mathrm{stoc}(t') \ketm = \frac{p(t)}{m} \frac{p(t')}{m} \left( \frac{ a }{ 2\pi \lambda } \right)^3 \frac{2 k_B T}{K_u} I(t-t'),
\]
and the inter-correlation between the force $F_\mathrm{stoc}(t)$ and the torque $T_\mathrm{stoc}(t')$ is found to be
\[
\bram F_\mathrm{stoc}(t) T_\mathrm{stoc}(t') \ketm = \dot{p}(t) \frac{p(t')}{m} \left( \frac{a}{2\pi \lambda} \right)^3 \frac{2 k_B T}{K_u} I(t-t').
\]
These correlations involve the same function $I(t-t')$ as before (\ref{eq:fonctionI}). From this study on the fluctuations of the spin waves, we conclude that the stochastic force $F_\mathrm{stoc}$ and the stochastic torque $T_\mathrm{stoc}$ behave as a multiplicative white noise on the domain wall.

\section{Reduction of domain wall width}

External forces, such as a magnetic field or an electrical current, lead to deformations of the domain wall profile. The most evident feature of such deformations is the creation of dipolar charges, which collectively build up the kinetic energy of the domain wall. Indeed the kinetic energy $ K_{\bot} \lambda^2 p^2 / 2 N_{dw} S^2 $ depends on the transverse anisotropy $K_{\bot}$, which in permalloy is due to the surface dipolar charges on the thin film surface. Another feature is that the width of the domain wall is reduced by the external magnetic field throughout its propagation~\cite{Thiaville2007}. In this section it is demonstrated that the latter feature does not specifically apply to the drag by a magnetic field but also appears in a similar manner under applied electrical currents.

The domain wall deformation is entirely described by the modes $c_{loc}(t)$, $c_{\kk}(t)$ and $c^{\dag}_{\kk}(t)$. In Section (\ref{interpretationBoundState}) the mode $c_\mathrm{loc}(t)$ has been interpeted as the kinetic energy of the domain wall. In the present section, we show that the modes $c_{\kk}^{\mathrm{def}}(t)$ and $c^{\dag \mathrm{def}}_{\kk}(t)$ can be interpreted as a reduction in the domain wall width, which subsequently appears as a change in the domain wall mass.

Departing from the Lagrangian (\ref{eq:LagrangienTotal}) and considering the Euler-Lagrange equation with respect to $c_{\kk}^{\dag}$, we obtain
\begin{eqnarray}
- \frac{S}{2i} \dot{c}_{\kk} &=& - \left( \hbar \Omega_{\kk} - S ( u -\dot{x}_0 ) \frac{k_x}{2} \right) c_{\kk} -  ( u -\dot{x}_0 ) v_{\kk} P \nonumber \\
&& + (u - \dot{x}_0) S \sum_{\kk \neq \mm} \; \left[ i v_{-\kk,\mm} ( c_{\mm} + c_{-\mm}^{\dag} ) \right. \nonumber \\
&& \left. - i v_{\mm,-\kk} ( c_{\mm} - c^{\dag}_{-\mm} ) \right] \, .
\label{eq:equationdiffck}
\end{eqnarray}
The mode $c_{\kk}$ has two components $c_{\kk}^{\mathrm{th}}$ and $c_{\kk}^{\mathrm{def}}$. The fast component $c^{\mathrm{th}}_{\kk}(t)$ ($\sim$ 4 GHz in permalloy) represents the thermal excitations, whereas the slow component $c_{\kk}^{\mathrm{def}}$ ($\sim$ 20 MHz for $d{x}_0/dt$=1m/s in permalloy) represents a deformation. As the differential equation (\ref{eq:equationdiffck}) is linear, the components $c_{\kk}^{\mathrm{th}}$ and $c_{\kk}^{\mathrm{def}}$ can be calculated independently. The slow component $c_{\kk}^{\mathrm{def}}$ is almost static $\dot{c}_{\kk}^{\mathrm{def}} = 0$ and corresponds to the ``static'' particular solution of (\ref{eq:equationdiffck}). In contrast the fast component of $c_{\kk}$ has no static part and is an homogeneous solution of (\ref{eq:equationdiffck}).

Let us calculate the deformation mode $c_{\kk}^{\mathrm{def}}$ from (\ref{eq:equationdiffck}) by approximating $\dot{c_{\kk}}^{\mathrm{def}} = 0$. It is useful to consider the Taylor expansion of ${c}_{\kk}^{\mathrm{def}}$ with respect to the relative velocity $u-\dot{x}_0$,
\begin{eqnarray}
c_{\kk}^{\mathrm{def}} = \sum_{n \geq 1} g_n (u - \dot{x}_0)^n \, .
\label{eq:TaylorCk}
\end{eqnarray}
Since the domain wall is much slower than the propagating spin waves $ (u-\dot{x}_0) / \lambda \ll \Omega_{\kk} $, it is sufficient to keep only the first order in the Taylor expansion (\ref{eq:TaylorCk}),
\begin{eqnarray}
c_{\kk}^\mathrm{def} \approx g_1 ( u - \dot{x}_0 ) \, .
\end{eqnarray}
On readily obtains
\begin{eqnarray}
c_{\kk}^{\mathrm{def}} \approx - \frac{ (u-\dot{x}_0) v_{\kk} }{ \hbar \Omega_{\kk} } p \, .
\end{eqnarray}
Therefore the deviation $\delta \phi^\mathrm{def}$ in the spherical angle $\phi$ due to deformation is
\begin{eqnarray}
\delta \phi^{\mathrm{def}} = \sum_{\kk} \left( c_{\kk}^{\mathrm{def}} + c_{-\kk}^{\mathrm{def}\ast} \right) \nu_{\kk}^{\phi} \xi_{\kk} \, , 
\end{eqnarray}
\begin{eqnarray}
\delta \phi^{\mathrm{def}}  &=& - 2 \sum_{\kk} \frac{ ( u -\dot{x}_0 ) p }{ \hbar \Omega_{\kk} } \left( \nu^{\phi}_{\kk} \right)^2 \frac{\pi}{2} \nonumber \\
&& \times \, \mathrm{sech} \left( \frac{k_x \lambda \pi}{2} \right)  \frac{1}{ \sqrt{ \omega_{\kk} N } } \xi_{\kk}  \, .
\end{eqnarray}
By letting $ \left( \nu^{\phi}_{\kk} \right)^2 = (1/4) \sqrt{ (\omega_{\kk} + \kappa)/\omega_{\kk}} $ and $ p = \left( \dot{x}_0 - u \right) m $, one finds
\begin{eqnarray}
\delta \phi^{\mathrm{def}} &=&  \frac{\pi}{4} \sum_{\kk} \frac{ S^2 ( u - \dot{x}_0 )^2 }{ K_u K_{\bot} \lambda } \frac{1}{\omega_{\kk}} \nonumber \\
&& \times \, \frac{ N_{\bot} }{a}  \mathrm{sech} \left( \frac{ k_x \lambda \pi }{ 2 } \right) \frac{1}{ \sqrt{ \omega_{\kk} N } } \xi_{\kk} \, .
\label{eq:deformationangle1}
\end{eqnarray} 

Let us now calculate the deformation angle $\delta \phi^{def}$ corresponding to a change $\delta \lambda$ in the domain wall width. Its comparison to (\ref{eq:deformationangle1}) will then allow us to estimate $\delta \lambda$. The $\phi$-angle describing the magnetization inside a Bloch domain wall is,
\begin{eqnarray}
\phi_0(x) = \pi + \mathrm{ cos }^{-1} \left[ \mathrm{ tanh } \left(  \frac{x-x_0}{\lambda_0}  \right) \right]
\end{eqnarray} 
If the domain wall width is modified by the spin current by $ \delta \lambda (u) $, the actual magnetization profile $\phi$ will deviate from the equilibrium Bloch profile $\phi_0$ as
\begin{eqnarray}
\phi &=& \phi_0 - \frac{ x - x_0 }{ \lambda^2 }   \sqrt{ 1 + \mathrm{ tanh }^2 \left(  \frac{x-x_0}{\lambda_0}  \right) } \; \delta \lambda \, .
\end{eqnarray}
The deformation angle $\delta \phi^{def} (\rr-\rr_0) = \phi(\rr) - \phi_{0}(\rr)$ can be expanded on the wave functions $\xi_{\kk}(\rr-\rr_0)$ as
\begin{eqnarray}
\delta \phi (\rr-\rr_0) &=& \sum_{\kk} p_{\kk}  \xi_{\kk}(\rr-\rr_0) \, ,
\end{eqnarray}
where
\begin{eqnarray}
p_{\kk} &=& \int \frac{d^3 r}{a^3} \delta \phi (\rr) \xi_{-\kk}(\rr) \nonumber \\
&=& - \int \frac{d^3 r}{a^3} \frac{ x - x_0 }{ \lambda^2 }   \sqrt{ 1 + \mathrm{ tanh }^2 \left(  \frac{x-x_0}{\lambda_0}  \right) } \; \delta \lambda \; \xi_{-\kk}(\rr) \, . \nonumber
\end{eqnarray}
Only the real part of the wave function $\xi_{-\kk}(\rr)$ will contribute to $p_{\kk}$,
\begin{eqnarray}
\Re e \left[ \xi_{-\kk}(\rr) \right] &=& \frac{1}{ \sqrt{ N \omega_{\kk} } } \left[ \cos \left[ k_x (x-x_0) \right] \mathrm{ tanh } \left( \frac{ x - x_0 }{ \lambda } \right) \right.\nonumber \\
&& \left. + k_x \lambda \sin k_x ( x - x_0 ) \right] \cos ( \kk_{\bot} . \rr ), \nonumber
\end{eqnarray}
accordingly,
\begin{eqnarray}
p_{\kk} = - \frac{ \delta \lambda }{ a } \frac{N_{\bot} \pi}{\sqrt{N \omega_{\kk}}}\, \mathrm{sech} \left( \frac{k_x \lambda \pi}{2} \right).
\label{eq:pdek}
\end{eqnarray}
Hence the change $\delta \lambda$ in the domain wall width corresponds to the deformation angle,
\begin{eqnarray}
\delta \phi^{\mathrm{def}} = -\sum_{\kk} \frac{ \delta \lambda }{ a } \frac{N_{\bot} \pi}{\sqrt{N \omega_{\kk}}}\, \mathrm{sech} \left( \frac{k_x \lambda \pi}{2} \right) \xi_{\kk}.
\label{eq:deformationangle2}
\end{eqnarray}
Comparing (\ref{eq:deformationangle1}) and (\ref{eq:deformationangle2}), and approximating $ \omega_{\kk} \approx 1 $, we finally obtain
\begin{eqnarray}
\frac{\delta \lambda}{\lambda} = - \frac{1}{4} \frac{ S^2 (u - \dot{x}_0)^2 }{ K_u K_{\bot} \lambda^2 }.
\label{eq:variationparoi}
\end{eqnarray}
Eq. (\ref{eq:variationparoi}) shows that the domain wall shrinks under the effect of spin transfer torque. According to (\ref{eq:DefinitionForceDeformation}), this reduction in width leads to a force $F_\mathrm{def}$ which is non-linear with respect to the relative velocity $u - \dot{x}_0$,   
\begin{eqnarray}
F_\mathrm{def} &=& - c \frac{d}{dt} \left[ p \frac{ S^2 (u-\dot{x}_0)^2 }{ K_u K_{\bot} \lambda^2 } \right],
\label{eq:ValeurForceDeformation}
\end{eqnarray}
with $ c = (\pi/32) \int^{+\infty}_{-\infty} d k_x \, \mathrm{sech}^2 \left(  \pi k_x/2 \right)  /( 1 + k^2_x ) \sim 0.1 $. In the case of permalloy nanowires, $K_{\bot}$ represents the demagnetizing energy and cancels with the factor $S^2$ in the numerator in (\ref{eq:ValeurForceDeformation}). Therefore $F_{\mathrm{def}}$ does not depend much on the transverse anisotropy $K_{\bot}$. 

\begin{figure}
	\begin{center}
		\includegraphics[scale=0.50]{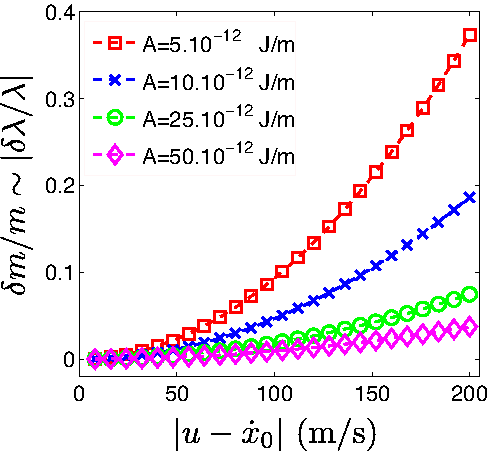}
	\end{center}
        \caption{Modification of the domain wall mass and of the domain wall width as a function of the relative velocity $|u-\dot{x}_0|$ and the exchange stiffness constant $A$. }
        \label{fig:DeformationMasseParoi}
\end{figure}

The deformation force $F_{\rm def}$ can be re-interpreted as a modification of the kinetic momentum and of the mass of the domain wall. As the domain wall deforms, its kinetic momentum becomes
\begin{eqnarray}
p + \delta p = p \left( 1 + \frac{ 3 c S^2 ( u - \dot{x}_0 )^2 }{ K_{\bot} A } \right),
\end{eqnarray}
and its mass is modified by
\begin{eqnarray}
\frac{ \delta m }{ m } = \frac{6c}{A \mu_0} (u - \dot{x}_0)^2 \left( \frac{\hbar}{g \mu_B} \right)^2.
\end{eqnarray}
The modifications of the domain wall mass and of the domain wall width are represented on Fig.\ref{fig:DeformationMasseParoi} as a function of the relative velocity $|u-\dot{x}_0|$ and for different values of the exchange stiffness constant $A$. 
They are almost identical $\delta m / m \sim - \delta \lambda / \lambda$.

\section{Temperature dependence of the domain wall mass}

The flow of an electrical current through a ferromagnetic wire modifies the spectrum of the magnons according to (\ref{eq:doppler}) and thereby shifts their average kinetic momentum to a finite value $ \bar{\kk} = \sum_{\kk} \kk n_{B}(\kk) / \sum_{\kk} n_{B}(\kk) $. The applied electrical current therefore leads to magnon current. Inside a domain wall, magnons behave slightly differently because of the singular coupling potential $ - S (u-\dot{x}_0) \sum_{\kk \mm} v_{\kk \mm} \hat{\theta}_{\kk} \hat{\phi}_{\mm} $ (\ref{eq:LagrangienTotal}). This potential may be interpreted as a modification of the domain wall's energy due to its interaction with the spin wave environment. This section is devoted to this coupling potential, whereby we investigate its consequences on the domain wall dynamics. Specifically, this potential will be shown to contribute to the forces $F_j$ (\ref{eq:forcej}) and $F_H$ (\ref{eq:forcem}) on the domain wall. Some light will then be shed on the process suggested in the Introduction (Section~\ref{Introduction}).

The relationship between the forces $F_\mathrm{j}$ and $F_\mathrm{H}$ and the statistics of the magnons will be established in Section~\ref{forces}. The force $F_\mathrm{j}$ arising from the dc-component of $u$ will be carried out in Section~\ref{adiabatic} and the force $F_\mathrm{H}$ arising from the ac-component $u(t)$ will then be calculated in Section~\ref{nonadiabatic}. Both these forces will be re-interpreted as a modification of the domain wall's effective mass. The domain wall mass will in turn become sensitive to the actual temperature of the ferromagnetic wire.

\subsection{\label{forces}Forces $F_\mathrm{j}$ and $F_\mathrm{H}$}

While the force $ F_\mathrm{j} $ depends on the correlation $\bram \delta \phi \delta \theta \ketm$ between $ \delta \theta $ and $ \delta \phi $, the force $ F_\mathrm{H} $ depends on the autocorrelation $ \bram \delta \theta^2 \ketm $ and $\bram \delta \phi^2 \ketm$. These correlations, which are finite under the effect of an electrical current, can be calculated with perturbation or linear response theory. For the sake of simplifying the notation, we will employ in this section a $2 \times 2$ matrix representation which is detailed in the Appendix.

The correlations between the angular deviations $\delta \phi$ and $\delta \theta$ are well described by the lesser magnon Green function $ {\cal{D}}^{<}(\rr,t,\rr',t') $ defined as
$$
i\hbar {\cal{D}}^{<} =
\left(
\begin{array}{cc}
\bram \delta \theta(\rr',t') \delta \theta(\rr,t) \ketm & \bram \delta \phi(\rr',t') \delta \theta(\rr,t) \ketm \\
\bram \delta \theta(\rr',t') \delta \phi(\rr,t) \ketm & \bram \delta \phi(\rr',t') \delta \phi(\rr,t) \ketm
\end{array}.
\right)
$$
The representation of the lesser function in the momentum space ${\cal{D}}^{<}(\kk,t,\kk',t')$ is obtained by expanding $\delta \theta$ and $\delta \phi$ on the wavefunctions $\xi_{\kk}$, according to (\ref{eq:expansionTheta}) and (\ref{eq:expansionPhi}). The representation in momentum space is quite useful as it reveals the relationship between the angular deviations $\delta \phi$,$\delta \theta$ and the statistics of the magnons. 

The force $F_\mathrm{H}$ depends on the diagonal component of the lesser function ${\cal{D}}^{<}_{\parallel} = {\cal{D}}^{<}_{\theta \theta} = {\cal{D}}^{<}_{\phi \phi}$,
\begin{eqnarray}
 F_\mathrm{H} = \sum_{ \kk \qq } f^{\mathrm{H}}( \kk,\qq ) i \hbar {\cal{ D }}^{<}_{\parallel} \left( \kk_{-},t , \kk_{+}, t \right),
\label{eq:forceFh}
\end{eqnarray}
where $\kk_{-} = \kk - \qq/2$ and $\kk + \qq/2$ and
\begin{eqnarray}
f^{\mathrm{H}}(\kk,\qq) &=& - \frac{K_u}{3\lambda} \frac{2\pi}{L} \frac{1}{i} \mathrm{sech} \left( \frac{\pi \lambda q_x}{2} \right) \lambda q_x   \\
&& \times \; \frac{ 1 + 3 (k_x \lambda)^2 + \frac{ (q_x \lambda)^2 }{ 4 } }{ \sqrt{ \omega_{\kk_-} \omega_{\kk_+} } } \left( \nu_{\kk_-}^{\theta} \nu_{\kk_+}^{\theta} + \nu_{\kk_-}^{\phi} \nu_{\kk_+}^{\phi}  \right). \nonumber
\end{eqnarray}
The force $F_\mathrm{j}$ depends on the off-diagonal component of the lesser function ${\cal{D}}_{\mathrm{off}}^{<}={\cal{D}}^<_{\phi \theta}$,
\begin{eqnarray}
F_\mathrm{j} = \sum_{\kk \qq} f^{\mathrm{j}}(\kk,\qq) \frac{d}{dt} i \hbar {\cal{D}}_{\mathrm{off}}^<(\kk_{-},t,\kk_{+},t),
\label{eq:forceFj}
\end{eqnarray} 
with
\begin{eqnarray}
f^\mathrm{j}(\kk,\qq) = - S v_{-\kk_{-},\kk_{+}}.
\end{eqnarray}
We recall that the coefficient $v_{-\kk_{-},\kk_{+}}$ was defined in (\ref{eq:definitionCoefficientVkm}) and that the coefficients $\nu_{\kk}^{\theta}$,$\nu_{\kk}^{\phi}$ were defined in (\ref{subsectionInteraction}).

As $f^\mathrm{H}(\kk,\qq)$ is odd in $\qq$ and even in $\kk$, the force $F_\mathrm{H}$ will be finite if ${\cal{ D }}^{<}_{\parallel} \left( \kk_{-},t , \kk_{+}, t \right)$ is odd in $\qq$ and even in $\kk$. It will be shown in Section~\ref{nonadiabatic} that the non-adiabatic response of the spin waves to a dynamical spin current $u(t)$ gives rise to such a lesser Green function ${\cal{D}}^<$. Besides as $f^\mathrm{j}(\kk,\qq)$ is odd with $\kk$ and even with $\qq$, $F_\mathrm{j}$ will be finite if  ${\cal{D}}_{\mathrm{off}}^<(\kk_{-},t,\kk_{+},t)$ is odd with $\kk$ and even with $\qq$. We will see in (\ref{adiabatic}) that this is the case for the adiabatic response of the spin waves to a dc spin current $u$.

\subsection{\label{adiabatic}Adiabatic response}

\begin{figure}
	\begin{center}
		\includegraphics[scale=0.4]{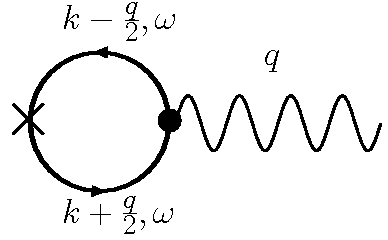}
	\end{center}
\caption{Representation of the force $ F_\mathrm{j} $ created by the spin waves when the relative velocity $ u - \dot{x}_0 $ is finite and the motion is adiabatic. The wavy line represents the magnon $q$ exchanged with the domain wall, the vertex $\times$ arises from the force $F_\mathrm{j}$ and the vertex $\bullet$ represents the potential ${\cal{V}}$.}
\label{fig:figure1}
\end{figure}

The process in Fig.~\ref{fig:figure1} depicts the adiabatic transfer of kinetic momentum between the propagating magnons and the domain wall, which arises from the coupling potential ${\cal{V}}$ (\ref{eq:interaction}). 
Due to adiabaticity, the energies of the incoming magnon $\kk - \frac{\qq}{2}$ and the outgoing magnon $\kk + \frac{\qq}{2}$ are identical.

\begin{figure}
	\begin{center}
		\includegraphics[scale=0.45]{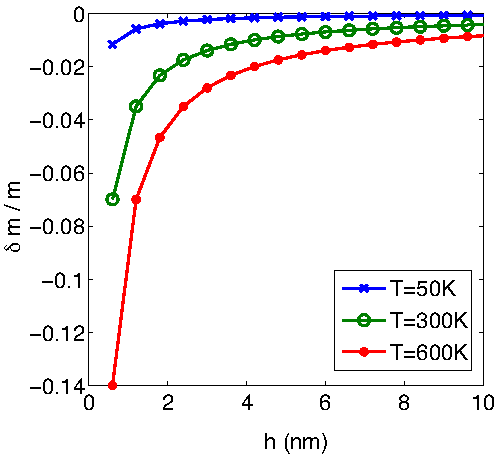}
	\end{center}
\caption{Modification of the domain wall's mass $\delta m / m$ as function of the wire's height $h$ and the temperature $T$, assuming $L_y=100$ nm and $\kappa=57$. }
\label{fig:ResponseAdiabatique1}
\end{figure}

The statistics of the magnons perturbed by the coupling potential ${\cal{V}}$ can be investigated by expanding the contour ordered magnon propagator ${\cal{D}}(\kk-\frac{\qq}{2},\kk+\frac{\qq}{2},\tau,\tau')$ up to the first order in ${\cal{V}}$,
\begin{eqnarray}
{\cal{D}}^{(1)}(\kk_{-},\tau,\kk_{+},\tau') = - S(u - \dot{x}_0)  \int_C d\tau_1  i v_{-\kk_{-},\kk_{+}}(\tau_1) \nonumber \\
 \times {\cal{D}}^{(0)}(\kk_{-},\tau-\tau_1) \sigma_y {\cal{D}}^{(0)}(\kk_{+},\tau_1-\tau'). \,\,
\label{eq:dysonlinear}
\end{eqnarray}
In (\ref{eq:dysonlinear}) the superscript ${(0)}$ represents equilibrium ($\dot{x}_0=u$), whereas the superscript ${(1)}$ represents first order perturbation theory. The lesser Green function ${\cal{D}}(\kk_{-},\kk_{+},t-t')$ can be obtained from the time ordered Green function with the Langreth formula $({\cal{D}} {\cal{D}} )^{<} = {\cal{D}}^r {\cal{D}}^< + {\cal{D}}^< {\cal{D}}^a$ and the fluctuation dissipation theorem,
\begin{eqnarray}
{\cal{D}}^{(0)<}(\kk,\omega) = n_B(\omega)  \left( {\cal{D}}^{(0)r}(\kk,\omega) - {\cal{D}}^{(0)a}(\kk,\omega) \right).
\label{eq:fluctudiss}
\end{eqnarray}
By noting $\mathrm{Im}(z)$ the imaginary part of $z$, and
$$
I =
\left( \begin{array}{cc}
1 & 0 \\
0 & 1
\end{array}
\right)
\;\;\; , \;\;\;
\sigma_y =
\left( \begin{array}{cc}
0 & -i \\
i & 0
\end{array}
\right) \, ,
$$
the lesser Green function is found in terms of the retarded (\ref{eq:magnonelectronr}) and advanced (\ref{eq:magnonelectrona}) propagators as
\begin{widetext}
\begin{eqnarray}
{\cal{D}}^{(1)<}(\kk_{-},\kk_{+},\omega) &=&  4 S (u-\dot{x}_0) v_{-\kk_{-}, \kk_{+}} n_B(\omega) \nonumber \\
&& \left[ I \; \mathrm{Im} \left( g_{+}^{(0)r}(\kk_{-},\omega) g_{+}^{(0)r}(\kk_{+},\omega) - g_{-}^{(0)r}(\kk_{-},\omega) g_{-}^{(0)r}(\kk_{+},\omega) \right) \right. \nonumber \\
&& + \left. \sigma_y \mathrm{Im} \left( g_{+}^{(0)r}(\kk_{-},\omega) g_{+}^{(0)r}(\kk_{+},\omega) + g_{-}^{(0)r}(\kk_{-},\omega) g_{-}^{(0)r}(\kk_{+},\omega) \right) \right] \, .
\label{eq:theoriePerturbation}
\end{eqnarray} 
\end{widetext}
From this equation, we can infer that ${\cal{D}}_{\parallel}^{(1)<}(\kk_{-},t,\kk_{+},t) = 0 $. Thus by inspection the adiabatic response of the spin waves to $(u-\dot{x}_0)$ doesn't contribute to the force $F_\mathrm{H}$ (\ref{eq:forceFh}). In contrast the off-diagonal component ${\cal{D}}^{(1)<}_{\mathrm{off}}(\kk_{-},\kk_{+},t)$ of the lesser Green function is finite when the domain wall velocity differs from the velocity of the spin current $ \dot{x}_0 \neq u $. It is obtained as
\begin{eqnarray}
i \hbar {\cal{D}}^{(1)<}_{\mathrm{off}}(\kk_{-},\kk_{+},t) = - 8 \pi S(u-\dot{x}_0) v_{-\kk_{-},\kk_{+}}  \left[ \frac{ \partial n_B }{\partial \epsilon} (\epsilon_{\kk}) \right], \nonumber
\end{eqnarray}
and leads to a force $F_\mathrm{j}$ (\ref{eq:forceFj})
\begin{eqnarray}
F_\mathrm{j} = - \frac{dp}{dt} \frac{ 8\pi K_{\bot} \lambda^2 }{ N_{dw} } \sum_{\kk ,\qq} \frac{\partial n_B(\epsilon_{\kk})}{\partial \epsilon}  \left( v_{-\kk_{-},\kk_{+}} \right)^2.
\end{eqnarray}
Being proportional to $dp/dt$, the force $F_\mathrm{j}$ may be re-interpreted as a modification of the domain wall's mass. In thin films, where the magnetization is uniform over the thickness of the wire, the change in the domain wall mass is found to be
\begin{eqnarray}
\frac{\delta m}{m} = \frac{k_B T a^2}{A} \frac{ 2\pi }{ N_z } \xi(\kappa,L_y) \, ,
\end{eqnarray}
where $N_z$ represents the number of atomic layers within the wire's thickness, $a$ is the lattice constant, $L_y$ stands for the width of the wire and $\xi(\kappa,L_y)$ is a function of $L_y$ and the anisotropies $\kappa=K_{\bot}/K_u$. The change in the mass $\delta m / m$ is presented in Fig.~\ref{fig:ResponseAdiabatique1} as a function of the wire thickness $h=N_z a$ and for different values of the temperature.
In a permalloy wire with a cross section 2 nm $\times$ 100 nm, the mass of the domain wall varies by $\sim$ 5\% between $T=0$ K and $T= 600$ K. This variation of the domain wall mass could be much larger near the Curie temperature $T_c \sim 750$ K.

\subsection{\label{nonadiabatic}Non-adiabatic response}

\begin{figure}[t]
	\begin{center}
		\includegraphics[scale=0.4]{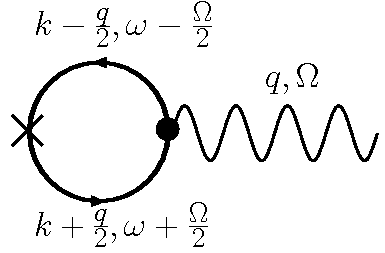}
	\end{center}
\caption{Representation of the force $ F_\mathrm{H} $ created by the spin waves when the excitation of the spin waves by the spin current is not adiabatic. The wavy line represents the momentum $\qq$ exchanged with the domain wall and the energy $\Omega$ transfered by the spin current, the vertex $\times$ arises from the force $F_\mathrm{H}$ and the vertex $\bullet$ represents the mixing potential ${\cal{V}}$.}
\label{fig:figure2}
\end{figure}

The process in Fig.~\ref{fig:figure2} shows the non-adiabatic transfer of kinetic momentum between the propagating magnons and the domain wall, which arises from the coupling potential ${\cal{V}}$ (\ref{eq:interaction}) when the domain wall is accelerating. 
Through this process, the energy of magnons is not conserved.

In the following, the domain wall motion is assumed to be harmonic $u(t) \propto e^{i \Omega t}$. This assumption will allow us to calculate the linear response of the spin waves to the domain wall acceleration in a more tractable way. The frequency $\Omega$ of the domain wall acceleration (e.g. $\sim$ 10 MHz) is much lower than the spin waves eigenfrequencies $\Omega_{\kk}$ ($\sim$ 5 GHz), but still induces an overlap between the different spin waves modes. As the actual overlap between the spin waves depend on their spectral linewidth, the lifetime $\tau_{\kk}$ of the magnons is a key physcial quantity in this problem. According to the phenomenological Gilbert damping, each magnon $\kk$ has a lifetime $\tau_{\kk}$ inversely proportional to its eigenfrequency and to the Gilbert coefficient, $\tau_{\kk}^{-1} = \alpha \Omega_{\kk}$. The linewidth of each spin wave mode is therefore about $ \Delta f =  0.01 \times 4$ GHz $\sim$ 40 MHz. As the wire length $L_x$ is generally much larger than the domain wall width $\lambda$, the energy gap between the peaks of two consecutive spin wave modes is of the order of $ \sqrt{K_uK_{\bot}}(2\pi  \lambda / L_x)^2 / 2 \sim 5 $ MHz.

\begin{figure}
	\begin{center}
		\includegraphics[scale=0.45]{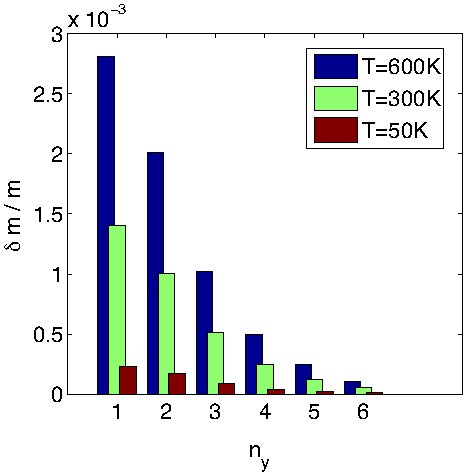}
	\end{center}
\caption{Contribution of the transverse spin wave modes $n_y$ to the change in the domain wall's mass $\delta m / m$. $L_y=400$ nm, $\alpha=0.01$ and $\kappa=K_{\bot}/K_u=57$.  }
\label{fig:ResponseNonAdiabatique1}
\end{figure}

For the sake of determining the linear response of the spin waves to the domain wall acceleration, we use the Dyson equation (\ref{eq:dysonlinear}) as a starting point. As $\Omega/\Omega_{\kk} \ll 1$, we linearize ${\cal{D}}^{(1)<}$ with respect to $\Omega$,
\begin{widetext}
\begin{eqnarray}
{\cal{D}}^{(1)<}(\kk-\frac{\qq}{2},\kk+\frac{\qq}{2},\omega,\omega+\Omega) = - i u_{-\kk+\frac{\qq}{2},\kk+\frac{\qq}{2}}(-\Omega) \left[ \frac{ \partial n_{B} }{ \partial \epsilon } (\omega) \right] \hbar \Omega    {\cal{D}}^{(0)r}(\kk-\frac{\qq}{2},\omega)  \sigma_y {\cal{D}}^{(0)a}(\kk+\frac{\qq}{2},\omega) \, .
\label{eq:GreenInfNonAd}
\end{eqnarray}
\end{widetext}
The diagonal component of the lesser magnon Green function is then obtained as,
\begin{eqnarray}
{\cal{D}}^{(1)<}_{\parallel}(\kk_{-},t,\kk_{+},t) = \frac{\alpha S}{m} \frac{ dp }{ dt } k_B T \Phi(k_x,q_x) \, ,
\label{eq:GreenNonAd}
\end{eqnarray} 
with
\begin{eqnarray}
 \Phi(k_x,q_x) = \frac{\lambda}{L} k_x q_x \lambda^2 \mathrm{csch} \left( \frac{ \pi \lambda q_x }{ 2 } \right) 4 \pi \frac{ \nu^{\theta}_{\kk_{-}} \nu^{\phi}_{\kk_{+}} }{ \sqrt{ \omega_{\kk_{-}} \omega_{\kk_{+}} } } \int \frac{d\omega}{\omega} \nonumber \\
\times \frac{ \epsilon_{\kk_{+}} - \epsilon_{\kk_{-}} }{ \left( ( \hbar \omega - \epsilon_{\kk_-} )^2 + (\alpha \epsilon_{\kk-})^2  \right)  \left( ( \hbar \omega - \epsilon_{\kk_+} )^2 + (\alpha \epsilon_{\kk_+})^2  \right)  } \, . \nonumber
\end{eqnarray}
The force $F_\mathrm{H}$ due to this non-adiabatic interaction is derived by carrying over the magnon Green function (\ref{eq:GreenNonAd}) to the definition of the force (\ref{eq:forceFh}). Because it is proportional to $dp/dt$, the force $F_\mathrm{H}$ can be re-interpreted as a modification of the domain wall's mass. The contributions of the transverse spin wave modes $n_y$ to $\delta m /m$ are shown on Fig.~\ref{fig:ResponseNonAdiabatique1} for different values of the temperature.
 The latter figure indicates that the change in the domain wall mass is in general less than $1/1000$. This implies that the non-adiabatic response of the spin waves to the domain wall acceleration does not affect domain wall motion significantly. This is in contrast with the adiabatic response of the spin waves (\ref{adiabatic}).

\section{Discussion and conclusion}

In conclusion, we have developed a theory of current-driven domain wall dynamics which includes spin-wave interactions. In general, a spin current $u$ traversing the wire will induce an interaction between the domain wall and the propagating spin wave modes. This interaction is proportional to the kinetic energy of the domain wall,  $m(u-\dot{x}_0)^2/2$. As this kinetic energy is governed by the difference between the spin current $u$ and the domain wall velocity $\dot{x}_0$, a domain wall moving at the spin current velocity will not interact with the spin waves. This coupling between the domain wall and the propagating magnons is shown to have two main effects on the domain wall dynamics: 1) Damping of the domain wall motion; 2) Renormalization of the domain wall effective mass. The damping accompanying the emission of magnons tends to decrease the energy of the domain wall thus tends to restore the solution, where the domain wall velocity and the spin current are identical $\dot{x}_0=u$. This dissipation channel is analogous to Landau dissipation in plasmas. The renormalization of the domain wall effective mass has two origins: the reduction in the domain wall width and the renormalization of the domain wall energy by thermal magnons.

In the presence of Gilbert damping and additional spin-wave interactions, our extended one-dimensional model of Bloch wall dynamics is given by
\begin{eqnarray}
 \dot{p} + \frac{ 2 K_{\bot} \m }{ S } \left( \alpha + \frac{\eta p^2}{ N_{dw} K_{\bot} \m } \right)  \dot{x}_0 = \nonumber \\
 \frac{ 2 K_{\bot} \m }{ S } \left( \beta + \frac{ \eta p^2 }{ N_{dw} K_{\bot} \m } u \right), \\ \label{eq:equationFinaleDyamique}
\frac{p}{\m}  =  \dot{x}_0 - u - \frac{ \alpha S }{ 2\m K_{\bot} } p.\label{eq:EquationFinaleCinematique}
\end{eqnarray}
The effective mass $\m$ takes into account the deformation of the domain wall and its coupling to the magnons. As such, it depends on the actual temperature of the wire. The emission of magnons, which is parametrized by the dimensionless coefficient $\eta$, is important if the coefficient $\eta$ is close to $1$. The equation of motion (\ref{eq:equationFinaleDyamique}) indicates that the emission of magnons ``contributes'' to the $\beta$ term. Since in (\ref{eq:equationFinaleDyamique}) the Gilbert damping coefficient $\alpha$ becomes shifted as $\alpha \rightarrow \alpha + {\eta p^2}/{ N_{dw} K_{\bot} \m } $, emission of spin waves also ``contributes'' to the damping. In the limit in which the only dissipation channel for the domain-wall walls is through spin-wave interactions, we find that the Galilean invariance in restored and $\alpha = \beta$.

The equations (\ref{eq:equationFinaleDyamique}) and (\ref{eq:EquationFinaleCinematique}) describe current induced domain wall motion by including some mechanisms that depend on the temperature. Temperature takes effect on domain wall dynamics through the domain wall mass $\m$ and also possibly through the emission coefficient $\eta$. According to Yamaguchi et al.\cite{Yamaguchi2007}, temperature inside the ferromagnetic nanowires may approach the Curie temperature $T_c$ because of Joule heating. According to our theory, such an increase in the temperature would appear as a renormalization of the domain wall mass. Our calculation has indicated a change in the domain wall mass of about 5\% between $T=600$K and $T=0$K. This change becomes even more important close to the Curie temperature $T_c$. Laufenberg et al.\cite{Laufenberg2006} found a significant decrease in the spin transfer efficiency, when the temperature was increased by few hundreds of Kelvin. These authors have suggested emission of magnons to be responsible for this loss of efficiency. According to our theory, emission of magnons would instead help to depin the domain walls. Thus the experimental results of Laufenberg et al. cannot be explained within our model.

\begin{figure}
	\begin{center}
		\includegraphics[scale=0.4]{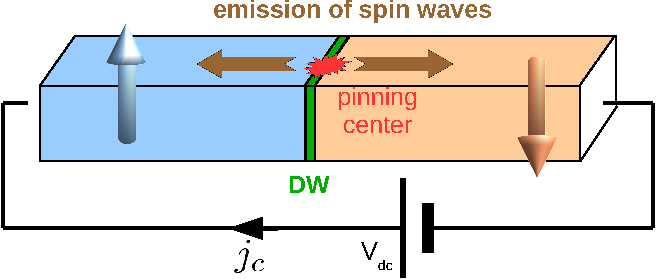}
	\end{center}
\caption{Proposed experiment for spin wave emission. A $dc$ power supply $V_{dc}$ generates a charge current $j_c$, which in turn excites a domain wall pinned inside a wire. As the spin current $u$ is finite but at the same time the domain wall is fixed, the relative velocity $u - \dot{x}_0$ is finite and spin waves are emitted.}
\label{fig:Experience1}
\end{figure}

According to the present theory, current-driven domain walls might be viewed as the generators of spin waves. One experimental realization of such spin-wave generation is proposed in Fig.~\ref{fig:Experience1}, where we show a domain wall pinned inside a wire that is traversed by a dc spin current. The action of the spin-current, as we have discussed in some detail above, acts to deform the domain wall and give rise to a non-equilibrium magnetic configuration. 
As its velocity vanishes $\dot{x}_0 =0$, its energy $\m(u-\dot{x}_0)^2/2$ will become very large. A part of this energy will then be dissipated to the medium through the emission of magnons.

\appendix

\section{Propagators}

\label{propagators}

The magnon propagators allow the linear response of the spin waves to be derived in a very tractable way. The magnon propagators are defined by means of the magnon operators $\hat{\phi}_{\kk}=\hat{c}_{\kk}+\hat{c}_{-\kk}^{\dag}$ and $\hat{\theta}_{\kk}=\frac{1}{i}\left( \hat{c}_{\kk} - \hat{c}_{-\kk}^{\dag} \right)$.
Specifically the time ordered magnon propagator ${\cal{D}}_{\alpha \beta}(\kk,\kk',t,t')$ for $\alpha,\beta \; \in \; \left\{ \theta , \phi  \right\} $, is defined as
\begin{eqnarray}
{\cal{D}}_{\alpha \beta}(\kk,\kk',t,t') =  -  \frac{i}{\hbar} \bra T \hat{\alpha}_{\kk}(t) \hat{\beta}_{-\kk'}(t') \ket  \, ,
\label{eq:definitionpropagator}
\end{eqnarray}
with
\[
T \hat{\alpha}_{\kk}(t) \hat{\beta}_{-\kk'}(t') = 
\left\{
\begin{array}{ll}
\hat{\alpha}_{\kk}(t) \hat{\beta}_{-\kk'}(t') & \; \mathrm{if} \;\; t > t'  \\ 
\hat{\beta}_{-\kk'}(t')  \hat{\alpha}_{\kk}(t) & \; \mathrm{if} \; \;  t < t' 
\end{array}
\right. \, .
\]
It is useful to let the times $t$ and $t'$ belong to a contour in the complex plane, which allows the Green's function (\ref{eq:definitionpropagator}) to contain information on the non equilibrium properties of the spin waves.
The lesser and greater propagators are defined as
\begin{eqnarray}
{\cal{D}}_{\alpha \beta}^< (\kk,\kk',t,t') &=& - \frac{i}{\hbar} \bra \hat{\beta}_{-\kk'}(t') \hat{\alpha}_{\kk}(t) \ket, \nonumber \\
{\cal{D}}_{\alpha \beta}^> (\kk,\kk',t,t') &=& - \frac{i}{\hbar} \bra \hat{\alpha}_{\kk}(t) \hat{\beta}_{-\kk'}(t') \ket, \nonumber 
\end{eqnarray}
and the retarded/advanced propagators are
\begin{eqnarray}
{\cal{D}}_{\alpha \beta}^{r}(t,t') &=& \theta(t-t') \left( {\cal{D}}_{\alpha \beta}^{>}(t,t') - {\cal{D}}_{\alpha \beta}^{<}(t,t') \right), \nonumber \\
{\cal{D}}_{\alpha \beta}^{a}(t,t') &=& \theta(t'-t) \left( {\cal{D}}_{\alpha \beta}^{<}(t,t') - {\cal{D}}_{\alpha \beta}^{>}(t,t') \right). \nonumber 
\end{eqnarray}
To simplify the notation, we represent the various magnon propagators by means of $2 \times 2$ matrices as
\[
{\cal{D}}(\kk,\kk',\tau,\tau') = 
\left(
\begin{array}{cc}
{\cal{D}}_{\theta,\theta}(\kk,\kk',\tau,\tau') & {\cal{D}}_{\theta,\phi}(\kk,\kk',\tau,\tau') \\
{\cal{D}}_{\phi,\theta}(\kk,\kk',\tau,\tau') & {\cal{D}}_{\phi,\phi}(\kk,\kk',\tau,\tau')
\end{array}
\right).
\]
We use the interaction picture with respect to the unperturbed spin-wave Hamiltonian ${\cal{H}}_0 = \sum_{\kk} \hbar \Omega_{\kk} c^{\dag}_{\kk} c_{\kk}$. The creation and annihilation operators of the magnons have the time dependence $\hat{c}_{\kk}(t) = \hat{c}_{\kk} e^{- i \Omega_{\kk}t}$, $\hat{c}_{\kk}^{\dag}(t) = \hat{c}_{\kk}^{\dag} e^{ i \Omega_{\kk}t}$.

Before calculating the response of the spin waves to the coupling potential ${\cal{V}}$, the unperturbed propagators need to be determined. The free retarded and the free advanced magnon propagators are obtained as
\begin{eqnarray}
{\cal{D}}^{(0)r}(\kk,\omega) = \left( \frac{ I + \sigma_y  }{ \hbar \omega - \epsilon_{\kk} + i0 } - \frac{ I - \sigma_y  }{ \hbar \omega + \epsilon_{\kk} + i0 }  \right) 
\label{eq:retarde}
\end{eqnarray}
and
\begin{eqnarray}
{\cal{D}}^{(0)a}(\kk,\omega) = \left( \frac{ I + \sigma_y  }{ \hbar \omega - \epsilon_{\kk} - i0 } - \frac{ I - \sigma_y  }{ \hbar \omega + \epsilon_{\kk} - i0 }  \right)  \, .
\label{eq:avance}
\end{eqnarray}
The free lesser Green function is given by the fluctuation-dissipation theorem,
\begin{eqnarray}
{\cal{D}}^{(0)<}(\kk,\omega) = n_B(\omega)  \left( {\cal{D}}^{(0)r}(\kk,\omega) - {\cal{D}}^{(0)a}(\kk,\omega) \right).
\end{eqnarray}
It turns out to be useful to introduce the electron-like propagators $g_{\pm}^{(0)r}(\kk,\omega)$ and $g_{\pm}^{(0)a}(\kk,\omega)$, defined as
\begin{eqnarray}
g_{\pm}^{(0)r}(\kk,\omega) &=& \frac{1}{ \hbar \omega \mp \epsilon_{\kk} + i \hbar \tau_{\kk}^{(-1)} } \, , \\
g_{\pm}^{(0)a}(\kk,\omega) &=& \left( g_{\pm}^{(0)r}(\kk,\omega) \right)^{\ast}.
\end{eqnarray}
These propagators are related to the magnon propagators (\ref{eq:retarde}) and (\ref{eq:avance}) through
\begin{eqnarray}
{\cal{D}}^{(0)r} &=& ( I + \sigma_y ) g_{+}^{(0)r} - ( I - \sigma_y ) g_{-}^{(0)r} 
\label{eq:magnonelectronr}
\end{eqnarray}
and
\begin{eqnarray}
{\cal{D}}^{(0)a} &=& ( I + \sigma_y ) g_{+}^{(0)a} - ( I - \sigma_y ) g_{-}^{(0)a}.
\label{eq:magnonelectrona}
\end{eqnarray}
The products of the retarded and advanced free magnon propagators, which appear throughout the calculation of the response, can be readily computed by using (\ref{eq:magnonelectrona}),(\ref{eq:magnonelectronr}), and by noticing that $(I+\sigma_y)(I-\sigma_y)=0$ and $(I\pm\sigma_y)^2=2(I\pm\sigma_y)$.

\begin{figure}
  	\begin{center}
  		\includegraphics[scale=0.55]{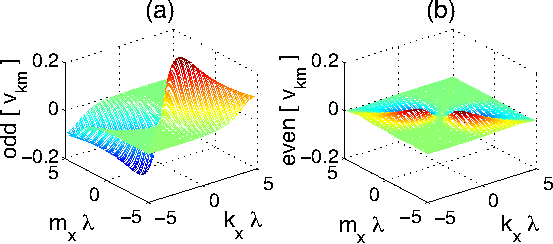}
  	\end{center}
          \caption{(a) Odd part $ (v_{\kk\mm} - v_{\mm\kk})/2 $ of coefficient $v_{\kk\mm}$. (b) Even part $ (v_{\kk\mm} + v_{\mm\kk})/2 $ of coefficient $v_{\kk\mm}$.  }
          \label{fig:oddPartVkm}
\end{figure}

The scattering potential ${\cal{V}}$ (\ref{eq:interaction}) may be divided into a ``linear'' term and a ``second order'' term with respect to the propagating modes. The ``linear'' term does not perturb the correlations of the spin waves since $\bram T \hat{c}_{\kk}^{\dag} \hat{c}_{\kk'} \hat{c}_{\kk'} \ketm = 0$ and $\bram T \hat{c}_{\kk}^{\dag} \hat{c}_{\kk'} \hat{c}_{\kk'}^{\dag} \ketm = 0$. However the second order term $ - S (u - \dot{x}_0) \sum_{\kk \mm} v_{\kk \mm} \hat{\theta}_{\mm} \hat{\phi}_{\kk} $ modifies these correlations. In the following we assume the coefficient $v_{\kk \mm}$ to be odd,
\begin{eqnarray}
v_{\kk \mm} = - v_{\mm \kk} \, .
\label{eq:ApproxParite}
\end{eqnarray}
Strictly speaking this assertion is wrong in general when $\kappa \neq 0$, however the plots in Fig.~\ref{fig:oddPartVkm} suggest that this is a good approximation.
In the interaction picture, the contour-ordered magnon propagator is given by 
\begin{eqnarray}
{\cal{D}}_{\alpha \beta}(\kk,\kk',\tau-\tau') &=&  -  \frac{i}{\hbar} \bra T \exp \left( - \frac{i}{\hbar} \int_C d\tau_1 {\cal{V}} (\tau_1) \right) \nonumber \\
&& \times \; \hat{\alpha}_{\kk}(\tau) \hat{\beta}_{-\kk'}(\tau') \ket  \, .
\end{eqnarray}
Linear response theory focuses on the first order expansion of ${\cal{D}}_{\alpha \beta}(\kk,\kk',\tau-\tau')$ with respect to the coupling potential ${\cal{V}}$. We use the superscript $^{(1)}$ to denote the first order expansion of the magnon propagator. Recalling Wick's theorem and using the relationship (\ref{eq:ApproxParite}), we readily obtain
\begin{eqnarray}
{\cal{D}}_{\alpha \beta}^{(1)}(\kk,\kk',\tau,\tau') = \int_C d\tau_1 u_{-\kk \kk'}(\tau_1) \left( {\cal{D}}_{\alpha \theta}^{(0)}(\kk,\tau-\tau_1) \right. \nonumber \\
\left. \times {\cal{D}}_{\phi \beta}^{(0)}(\kk',\tau_1-\tau')  - {\cal{D}}_{\alpha \phi}^{(0)}(\kk,\tau-\tau_1) {\cal{D}}_{\theta \beta}^{(0)}(\kk',\tau_1-\tau') \right) \, . \nonumber
\end{eqnarray}
Employing the $2 \times 2$ matrix notation, we finally find
\begin{eqnarray}
{\cal{D}}^{(1)}(\kk,\kk',\tau,\tau') = \int_C d\tau_1 i u_{-\kk \kk'}(\tau_1)   \\
 \times {\cal{D}}^{(0)}(\kk,\tau-\tau_1) \sigma_y {\cal{D}}^{(0)}(\kk',\tau_1-\tau') \, .
\label{eq:EquationDyson}
\end{eqnarray}
Equation (\ref{eq:EquationDyson}) expresses the linear response of the spin waves to the coupling potential ${\cal{V}}$ (\ref{eq:interaction}).

\begin{acknowledgments}
YLM thanks D. Ravelosona for the important guidance and supervision of his work during his Phd thesis. He gratefully acknowledges very fruitful discussions with J. Shibata, H. Kohno, S. Kasai, A. Takeuchi, K. Hosono, E.E. Fullerton, F. Pi\'echon, A. Thiaville, T. Devolder, C. Chappert, V. Vlaminck and T. Ono. He also thanks Tokyo Metropolitan University for financial support during his stay in Minami Osawa and support from the CILOMAG contract of the French National Research Agency (ANR).
\end{acknowledgments}

\bibliography{Biblio5}

\end{document}